\documentclass[prd,a4paper,showpacs,preprint,byrevtex]{revtex4}
\usepackage{graphicx}
\usepackage{dcolumn}
\usepackage{amsmath}
\usepackage{array}
\usepackage{bm}
\usepackage{textcomp}
\def\mathun{\hbox{ 1\hskip -3pt l}}
\begin{document}
\title{Short-range potentials from QCD at order $g^2$}

\author{Vincent \surname{Mathieu}}
\thanks{IISN Scientific Research Worker}
\email[E-mail: ]{vincent.mathieu@umh.ac.be}
\author{Fabien \surname{Buisseret}}
\thanks{FNRS Research Fellow}
\email[E-mail: ]{fabien.buisseret@umh.ac.be}
\affiliation{Groupe de Physique Nucl\'{e}aire Th\'{e}orique,
Universit\'{e} de Mons-Hainaut,
Acad\'{e}mie universitaire Wallonie-Bruxelles,
Place du Parc 20, BE-7000 Mons, Belgium}

\date{\today}
\begin{abstract}
We systematically compute the effective short-range potentials arising from second order QCD-diagrams related to bound states of
quarks, antiquarks, and gluons. Our formalism relies on the assumption that the exchanged gluons are massless, while the
constituent gluons as well as the lightest quarks acquire a nonvanishing constituent mass because of confinement. We recover the well-known Fermi-Breit interaction in the heavy quark limit. Such an effective potential has been proved in the past to be relevant for the building of accurate potential models describing usual hadrons (mesons and baryons). The general potentials we obtain in this work are also expected to be useful in the understanding of exotic hadrons like glueballs and $q\bar q g$ hybrids. In particular, we compute for the first time an effective short-range quark-gluon potential, and show the existence of a quadrupolar interaction term in this case. We also discuss the influence of a possible nonzero mass for the exchanged gluons.
\end{abstract}

\pacs{12.39.Ki: Relativistic quark model; 12.39.Pn: Potential model}

\maketitle

\section{Introduction}

\par Mesons and baryons have been intensively studied in the framework of potential models for a long time \cite{old,old2}. Moreover, since QCD is a nonabelian gauge theory, the gluons are able to form a bound state, at least in principle. These bound states of two, three or more gluons are called glueballs. The theoretical computation of the glueball spectrum is a problem which currently deserves much interest, either in lattice QCD~\cite{lat}, or within the framework of effective models~\cite{corn,bar,simo99,hou,Brau04,Math}. Other so-called ``exotic" hadrons are hybrid mesons: Mesons in which the color field is in an excited state. They can be seen as a three-body $q\bar q g$ state, with a constituent gluon. Again, hybrid mesons are the subject of many works, in lattice QCD~\cite{lath,scabicu}, and effective models \cite{Mand,fabhyb,new,cg1,cg2}. The experimental detection of exotic hadrons is also an active field of research, although no definitive conclusion has been drawn \cite{klempt}. The confrontation between theoretical approaches and experimental data appears thus particularly interesting in order to confirm or invalidate some exotic candidates.
\par Let us now turn our attention to effective QCD models. It is a well-known fact that the short-range interaction between two heavy quarks is given by the Fermi-Breit potential \cite[p.~395]{land}. This corresponds to the one-gluon exchange process between the quarks, and contains fine structure terms like spin-orbit interaction and tensor force for example. Effective models with a linear confining potential and the Fermi-Breit potential have been showed to correctly describe the meson and baryon spectrum for a long time \cite{old}. Apart from phenomenological arguments, the static potential between a quark and an antiquark has also been intensively studied in numerous theoretical works. At large distances, typically for distances larger than $2.5$ GeV$^{-1}$, the static potential indeed appears to be linearly rising (see for example the lattice calculations of ref.~\cite{weis}). At short distances however, the static potential has been shown to be no longer linear and nonperturbative but rather perturbative and Coulombic, just as is the static limit of the Fermi-Breit interaction~\cite{pineda}. 

Logically, the same kind of models -- linear confinement and Coulomb term at long- and short-range respectively -- has since been applied to describe exotic hadrons, like glueballs \cite{corn,bar,simo99,hou,Brau04,Math} and hybrids \cite{Mand,fabhyb,new}. However, if the derivation of the Fermi-Breit potential between two quarks is a standard procedure, it is not the case for the interactions between two gluons (see for example the discrepancies between refs.~\cite{corn,bar,hou}). Moreover, to our knowledge, the equivalent of the Fermi-Breit potential between a quark and a gluon has not yet been computed. Since recent references, based on lattice QCD computations, validate the picture of a glueball as a bound state of constituent gluons~\cite{glue2} and of a hybrid meson as a $q\bar q g$ system~\cite{scabicu}, the accurate knowledge of the gluon-gluon and quark-gluon potentials appears to be crucial point to further understand the properties of these exotic hadrons. It is thus of interest to compute all these potentials in a systematic way, within a single formalism, in order to obtain a set of potentials which should be applicable to most hadrons, either exotic or not. These computations are performed in this work.
\par Our paper is organized as follows. In sec.~\ref{gene}, we introduce the general procedure for extracting the effective potential corresponding to a particular matrix element. Before applying it to all the possible second order QCD-diagrams related to bound states, we have to know the non relativistic wave functions of the \textit{in} and \textit{out} states. This is presented in sec.~\ref{inout}. Then, we compute the effective potentials for the diagrams involving pairs of quarks and antiquarks in sec.~\ref{qqint}. In sec.~\ref{ggint}, we consider the interactions between two gluons, and the interactions between (anti)quarks and gluons are treated in sec.~\ref{qgint}. The exchanged gluons are massless in our formalism. Since it is not the case in some approaches (see refs.~\cite{corn,hou}), we then discuss in sec.~\ref{applic} the influence of a possible nonzero mass for the exchanged gluons. In sec.~\ref{valid}, we discuss the validity of the potentials which are computed. Since we want our potentials to be relevant in various systems -- exotic or non exotic hadrons --, we compute in sec.~\ref{colfact} the color factors appearing in the various potentials we obtain. Finally, we draw some conclusions in sec.~\ref{conclu}. We also added two appendices, where we recall the Feynman rules for QCD (appendix~\ref{feyn}) and some useful Fourier transforms (appendix~\ref{fourier}).

\section{General method}\label{gene}

Let us consider a general Feynman diagram of order $g^2$, with $g$ the strong coupling constant
\begin{figure}[ht]
  \begin{center}
    \includegraphics*[width=3.0cm]{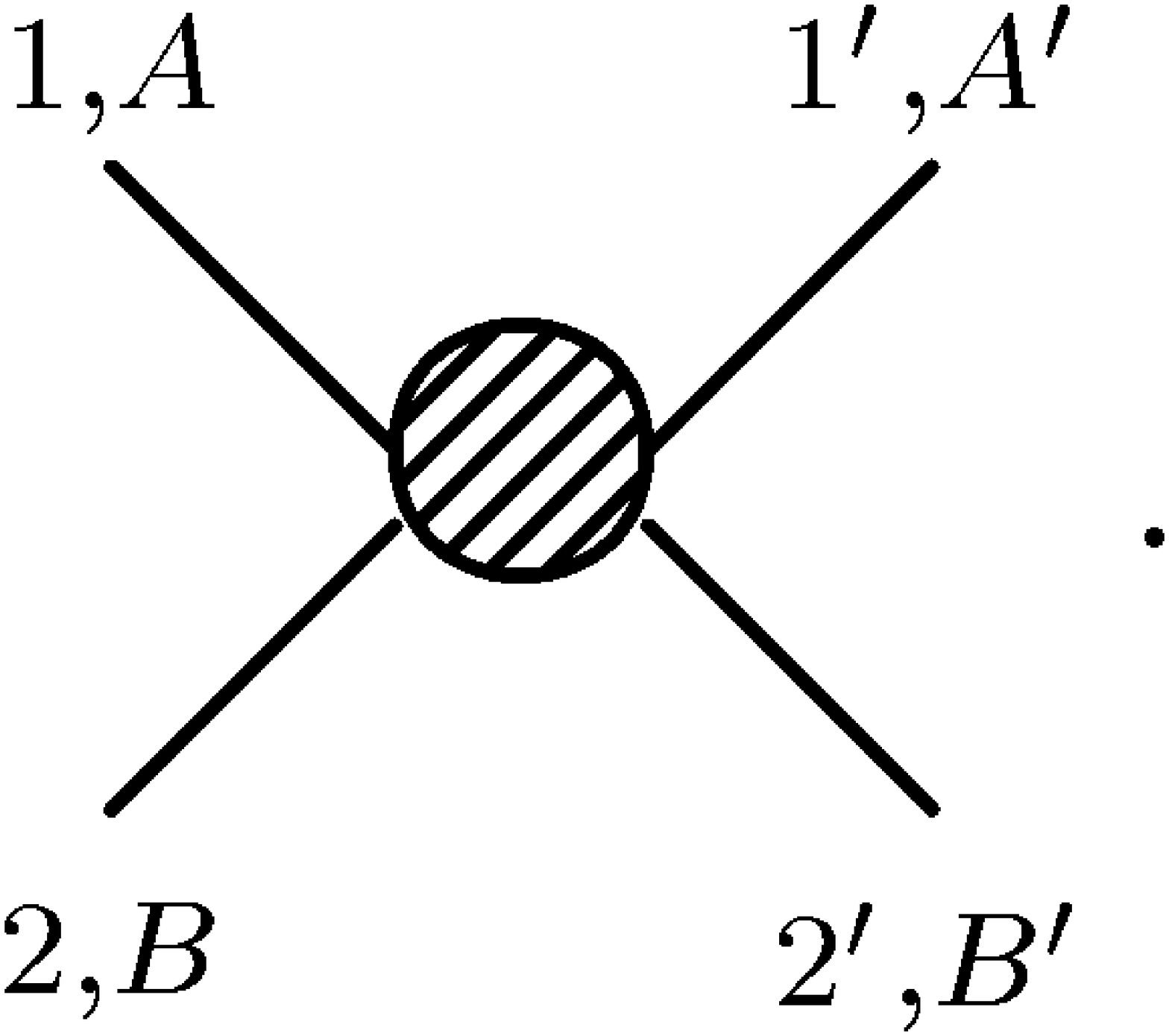}
  \end{center}
\end{figure}
\\
$A,B,A',B'$ denote the color indices of the particles $1$, $2$, $1'$, and $2'$. As the goal of this work is to compute effective potentials for bound states of quarks, antiquarks, or gluons, we will restrict ourselves to diagrams where the initial and final particle content are the same: $1=1'$ and $2=2'$. The application of Feynman rules in momentum space, summed up in Appendix~\ref{feyn}, leads us formally to the following matrix element (in units where $\hbar=1$)
\begin{equation}\label{mif0}
    M_{fi}=-i(2\pi)^4 \delta^4(P-P')\, M(\Psi^C_{{\rm i},\, \alpha};\Psi^{C'}_{{\rm o},\, \beta};q),
\end{equation}
where $q$ is the exchanged $4$-momentum, and $\Psi^C_{{\rm i},\alpha},\, \Psi^{C'}_{{\rm o},\beta}$ are the properly normalized wave functions of the \textit{in} and \textit{out} states. In our notations, $\alpha,\beta=1,2$ label the particles, $C=A,B$, and $C'=A',B'$. These wave functions can be schematically decomposed as
\begin{equation}\label{fodef}
    \Psi^C=\phi^C\otimes\psi(p),
\end{equation}
 with $\phi^C$ the color wave function, and $\psi(p)$ the spatial wave function in momentum space, $p$ being the $4$-momentum of the particle. The explicit form of these wave functions in the nonrelativistic limit will be given in sec.~\ref{inout}. Then, the crucial and most complex step of the procedure is to rewrite $M$ as
\begin{equation}\label{mif}
M=\Psi^{A'}_{{\rm o},1}\Psi^{B'}_{{\rm o},2}\left[g^2 {\cal O}_{A'B'AB} U(q^0,\bm q,\bm p_\alpha,\bm{S}_\alpha)\right]\Psi^A_{{\rm i},1}\Psi^B_{{\rm i},2}.
\end{equation}
$\bm{p}_\alpha$ and $\bm{S}_\alpha$ are the momentum and spin of the particle $\alpha$. In the latter, bold quantities will always denote a vector. The spin operators appear through the rearrangement of the wave functions, and ${\cal O}_{A'B'AB}$ is a color operator emerging from the vertices. We will show in sec.~\ref{colfact} that this operator turns out to be a real number when correctly contracted with the color functions of the particles, $\phi^C$. We denote ${\cal C}$ this color factor.
\par Once the decomposition~(\ref{mif}) is known, the effective potential we want to find is readily obtained in position space by a Fourier transform, denoted as ${\cal F}$, which changes $\bm{q}$ in $\bm{r}$, the relative position of the two interacting particles:
\begin{equation}\label{pot_four}
    V(\bm{r})={\cal C}\alpha_S\ {\cal F}\left[4\pi U(q^0,\bm{q},\bm{p}_\beta,\bm{S}_\beta)\right].
\end{equation}
We used the definition $g^2=4\pi\alpha_S$, which has the advantage of simplifying the Fourier transforms. All the Fourier transforms we need are listed in Appendix~\ref{fourier}.
\par We have at last to make a remark which will be valid in all our work: The effective potentials, and all the other quantities appearing in the latter, will be computed at the order $c^{-2}$, with $c$ the speed of light. This is a standard approximation scheme, justified because the most relevant contributions as spin-orbit, tensor interaction, \dots, arise from these first relativistic corrections.

\section{\textit{In} and \textit{Out} states}\label{inout}
\subsection{Quarks and antiquarks}\label{qwf}
Quarks and antiquarks are spin-$1/2$ particles, so they are described by Dirac spinors, denoted as $u^i(\bm p)$ for the quarks and $v^j(\bm p)$ for the antiquarks, with $\bm p$ the momentum of the particle and $i,j$ their color indices. The wave function of a Dirac spinor, whose mass is $m$, can be computed at the order $c^{-2}$ thanks to a Foldy-Wouthuysen transformation \cite{FW}. In our notations, it can be recast in the form~(\ref{fodef}), and reads
\begin{eqnarray}\label{uvidef}
    u^i(\bm p) &=& \phi^i\otimes u(\bm p),\nonumber \\
    v^j(\bm p) &=&  \phi^j\otimes v(\bm p),
\end{eqnarray}
with $\phi^k$ the color functions, and
\begin{eqnarray}\label{uvdef}
    u(\bm p) &=& \sqrt{2m}\ \begin{pmatrix}(1-\bm p^2/8m^2c^2)\, w\\
                            (   \bm \sigma\cdot\bm p/2mc)\ w\end{pmatrix},\nonumber\\
    v(\bm p) &=& \sqrt{2m}\ \begin{pmatrix}(\bm \sigma\cdot\bm p/2mc)\ w\\
                                (1-\bm p^2/8m^2c^2)\, w\end{pmatrix},
\end{eqnarray}
where $w$ is a two component spinor and ${\bm \sigma}$ are the Pauli matrices (see Appendix \ref{feyn}). In order to rewrite the matrix element on the form (\ref{mif}), we will have to isolate a factor $\sqrt{2m}\, w$ for each wave function.

\par The following useful relations hold for both $u(\bm p)$ and $v(\bm p)$ \cite[p. 281]{land}
\begin{subequations}\label{rel_q}
\begin{eqnarray}
    \bar u(\bm p')\gamma^0u(\bm p) &=& 2m w^\dag\left(1-\frac{\bm q^2 + 2i\bm\sigma\cdot(\bm q\times\bm p)}{8m^2c^2}\right)w,\\
    \bar u(\bm p')\bm\gamma u(\bm p) &=& 2m\, w^\dag\left(\frac{i\bm\sigma\times\bm q}{2mc}+\frac{\bm q+2\bm p}{2mc}\right)w,
\end{eqnarray}
\end{subequations}
with $\bm q=\bm p'-\bm p$, $\bar u=u^\dag\gamma^0$.
\par An important relation between the Pauli matrices is 
\begin{equation}
    (\bm\sigma\bm A)(\bm\sigma\bm B) = \bm A\cdot\bm B+i\bm\sigma\cdot(\bm A\times\bm B),
\end{equation}
in which $\bm A$ and $\bm B$ are two vectorial operators. The spin operator of a quark or an antiquark is defined by $\bm S=\bm\sigma/2$.
\subsection{Gluons}\label{gwf}

A massive, colored, spin-$1$ particle is described by the wave function
\begin{equation}
    \epsilon^a_\nu(\bm p)=\phi^a \otimes\epsilon_\nu(\bm p),
\end{equation}
where $\epsilon_\mu$ is a polarisation vector satisfying the Proca equation with the transverse condition
\begin{equation}\label{gluc}
p^\nu\epsilon_\nu(\bm p)=0.
\end{equation}
Relation~(\ref{gluc}) together with the normalisation condition $\epsilon^\nu\epsilon_\nu=-1$ leads to the nonrelativistic expression
\begin{eqnarray}\label{epsdef1}
    \epsilon(\bm p) &=& \left(\frac{\bm\epsilon\cdot\bm p}{mc},\bm\epsilon+\frac{\bm\epsilon\cdot\bm p}{2m^2c^2}\bm p\right),
\end{eqnarray}
with
\begin{equation}
    p=\left(mc+\frac{\bm p^2}{2mc}, \bm p\right),\ \ \rm{and}\ \ \bm \epsilon^2=1,
\end{equation}
in agreement with refs.~\cite{bar,hou}. The polarization vector $\bm\epsilon$ does not depend on the momentum $\bm p$ anymore. The factor to isolate in the matrix element will now be given by $\sqrt{2p^0}\,  \bm \epsilon$, by analogy with the Dirac spinors.  The spin operators for a spin-$1$ particle are defined by the matrices $\bm S$, where
\begin{equation}\label{spin1def}
(S^k)^{lm}=-i\varepsilon^{klm}.
\end{equation}
 $\varepsilon^{klm}$ is the Levi-Civita symbol.
The following useful relation involves the spin-$1$ matrices
\begin{equation}\label{spin1prop}
    A_iB_j=\left[\bm A\cdot\bm B - (\bm S\cdot\bm B)(\bm S\cdot\bm A)\right]_{ij}.
\end{equation}

\par For QCD applications as those we consider here, the spin-$1$ particles which have to be taken into account are the gluons. These are massless particles, so, one can wonder if formula (\ref{epsdef1}) makes sense. In our framework however, exchanged gluons, which are true massless gauge particles, have to be distinguished from the constituent gluons of the $in$ and $out$ states. These ones are confined inside a hadron, and one can consider that even if the bare gluon mass is zero, they acquire a nonzero constituent mass due to the confining interaction. $m$ has thus here to be understood as the constituent gluon mass in a particular bound state, which typically has a value around $0.5-0.6$ GeV~\cite{new,silva}. This argument has already been used in refs.~\cite{bar,Mand}, and will be further developed in sec.~\ref{bfapp}. Let us note that the same argument holds in the case of the massless $u$ and $d$ quarks, whose constituent mass is around $0.3$ GeV~\cite{new}.

\section{Interactions between quarks and antiquarks}\label{qqint}

\subsection{Fermi-Breit potential}\label{fbpot}
As an illustration of the general method presented in the previous section, we begin by the best-known diagram, that is the one-gluon exchange between two quarks. Applying the Feynman rules of Appendix~\ref{feyn}, we can write the matrix element corresponding to the diagram

\begin{figure}[h]
  \begin{center}
    \includegraphics*[width=3cm]{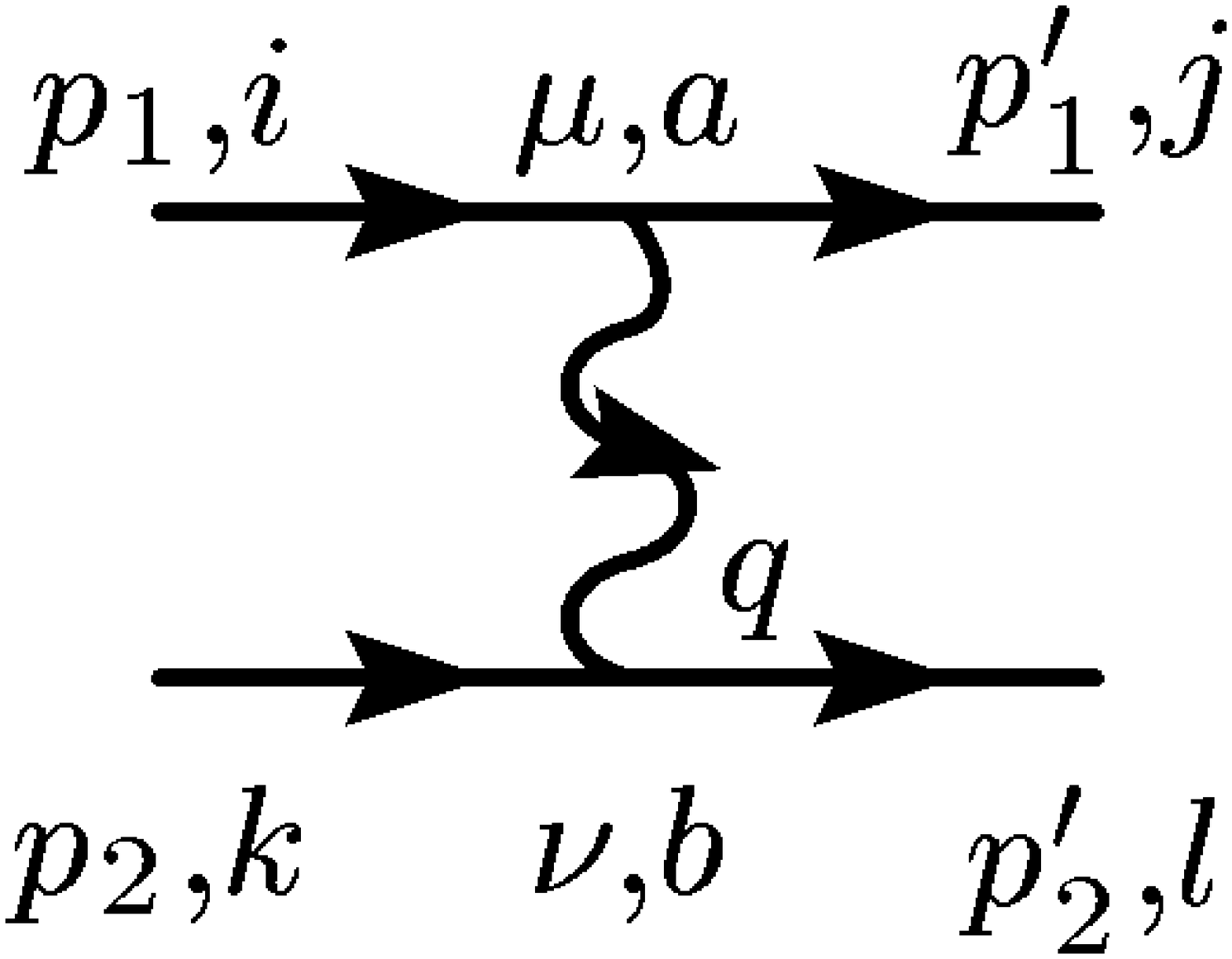}
  \end{center}
\end{figure}

\noindent in the form~(\ref{mif0}), which is
\begin{equation}\label{qqstep1}
    M_{fi}=-i(2\pi)^4\delta^4(P'-P) (-g^2)  {\cal O}_{1;ijkl}\,
    J^\mu_1(q) D_{\mu\nu}J^\nu_2(- q).
\end{equation}
The color operator is given by
\begin{equation}
    {\cal O}_{1;ijkl}=\frac{\lambda^a_{ji}}{2} \frac{\lambda_{a,lk}}{2},
\end{equation}
and the conserved current of the quarks by
\begin{equation}\label{Fmu}
    J^\mu_\alpha(q)=\bar u'_\alpha(p+q)\, \gamma^\mu\,  u_\alpha(p).
\end{equation}
We omitted to note the color functions in order to simplify the notations. The explicit computation of the color factors will be done in sec.~\ref{colfact}. Let us call ${\cal C}_1$ the color factor corresponding to ${\cal O}_{1;ijkl}$.
\par As the current $J^\mu_\alpha(\bm q)$ is obviously known thanks to the relations~(\ref{rel_q}), we only need to specify the form of the propagator $D_{\mu\nu}$ in order to develop eq.~(\ref{qqstep1}).  Let us note that
\begin{equation}
    \bm q= \bm p'_2-\bm p_2=\bm p_1-\bm p'_1.
\end{equation}
An interesting gauge choice is then to set $\theta=0$ in the propagator~(\ref{Ddef}). This leads to the expressions
\begin{equation}\label{propadef}
    D_{00}=-\frac{1}{\bm q^2},\quad D_{0i}=0,\quad D_{ik}=\frac{1}{\bm q^2}\left(\delta_{ik}-\frac{q_i q_k}{\bm q^2}\right).
\end{equation}
Putting all these elements together, one can rewrite eq.~(\ref{qqstep1}) in the form (\ref{mif}), i. e.
\begin{equation}\label{qqstep2}
    M=(4m_1m_2)\,  w'^\dagger_1 w'^\dagger_2\, \left[g^2\, {\cal C}_1\, U(\bm q,\bm p_1,\bm p_2)\right]w_1w_2,
\end{equation}
with
\begin{eqnarray}\label{eq22}
U(\bm q,\bm p_1,\bm p_2)&=&\frac{1}{\bm q^2}-\frac{1}{8m^2_1 c^2}-\frac{1}{8m^2_2 c^2}-\frac{i\bm\sigma_1\cdot(\bm q\times\bm p_1)}{4m^2_1c^2 \bm q^2}+\frac{i\bm\sigma_2\cdot(\bm q\times\bm p_2)}{4m^2_2c^2 \bm q^2}\nonumber \\
&&+\frac{(\bm p_1\cdot\bm q)(\bm p_2\cdot\bm q)}{m_1 m_2\bm q^4 c^2}
-\frac{\bm p_1\cdot\bm p_2}{m_1 m_2\bm q^2 c^2}+\frac{i\bm\sigma_1\cdot(\bm q\times\bm p_2)}{2 m_1m_2c^2 \bm q^2}-\frac{i\bm\sigma_2\cdot(\bm q\times\bm p_1)}{2 m_1m_2c^2 \bm q^2}\nonumber\\
&&+\frac{(\bm\sigma_1\cdot\bm q)(\bm\sigma_2\cdot\bm q)}{4m_1m_2c^2\bm q^2}-\frac{\bm\sigma_1\cdot\bm\sigma_2}{4m_1m_2c^2}.
\end{eqnarray}
In the latter, the $c$ factors will not be written anymore in order to simplify the notations.
The last step is to compute the Fourier transform of $U$, as formulated in eq.~(\ref{pot_four}). By using the relations of Appendix~\ref{fourier}, we obtain the following effective potential
\begin{eqnarray}\label{pot_1}
    V(\bm r)&=&{\cal C}_1 \alpha_S\left\{\frac{1}{r}-\frac{\pi}{2}\left(\frac{1}{m^2_1}+\frac{1}{m^2_2}\right)\delta^3(\bm r)-\frac{\bm L_1\cdot\bm S_1}{2m^2_1r^3}+\frac{\bm L_2\cdot\bm S_2}{2m^2_2r^3}\right.\nonumber\\
    &&-\frac{1}{2m_1m_2r}\left[\bm p_1\cdot\bm p_2+\frac{(\bm p_1\cdot\bm r)(\bm p_2\cdot\bm r)}{r^2}\right]-\frac{1}{m_1m_2r^3}\left[\bm L_1\cdot\bm S_2-\bm L_2\cdot\bm S_1\right]\nonumber\\
    &&+\frac{1}{m_1m_2r^3}\left[\bm S_1\cdot\bm S_2-3\frac{(\bm S_1\cdot\bm r)(\bm S_2\cdot\bm r)}{r^2}\right]-\left.\frac{8\pi}{3m_1m_2}\bm S_1\cdot\bm S_2\delta^3(\bm r)\right\},
\end{eqnarray}
where $\bm L_i=\bm r\times\bm p_i$. This is the well-known Fermi-Breit interaction \cite[p. 395]{land}. It is interesting to rewrite this contribution by using 
\begin{subequations}\label{change}
\begin{equation}
    \bm L=\frac{m_2\bm L_1 - m_1 \bm L_2}{m_1+m_2}\, ,\quad \bm \Omega=\bm L_1+\bm L_2\, ,
\end{equation}
which are simply the relative and center of mass variables of the interacting pair, and also by introducing new spin variables as follows
\begin{equation}
\bm S=\bm S_1+\bm S_2,\quad	\bm \delta=\bm S_1-\bm S_2\, .
\end{equation}
Potential (\ref{pot_1}) then becomes
\end{subequations}

\begin{eqnarray}\label{pot_1bis}
    V(\bm r)&=&{\cal C}_1 \alpha_S\left\{\frac{1}{r}+\pi\left(\frac{6-4\bm S^2}{3m_1m_2}-\frac{1}{2m^2_1}-\frac{1}{2m^2_2}\right)\delta^3(\bm r)-\frac{1}{2m_1m_2}\left[\frac{\bm p_1\cdot\bm p_2}{r}+\frac{(\bm p_1\cdot\bm r)(\bm p_2\cdot\bm r)}{r^3}\right]\right.\nonumber\\
    &&-\left(\frac{m^2_2+m^2_1+4m_1m_2}{4m_1^2m_2^2}\right)\frac{\bm L\cdot\bm S }{r^3}+\frac{1}{4m_1m_2}\frac{\bm \delta\cdot\bm \Omega}{r^3}+
    \left(\frac{m_1^2-m_2^2}{4m_2^2m_1^2}\right)\frac{\bm L\cdot\bm \delta}{r^3}\nonumber\\
    &&+\frac{(m_2-m_1)}{4(m_1+m_2)m_1m_2}\frac{\bm S\cdot\bm \Omega}{r^3}+\left.\frac{1}{2m_1m_2}\left[\frac{\bm S^2}{r^3}-3\frac{(\bm S\cdot\bm r)^2}{r^5}\right]\right\}.
\end{eqnarray}
It can be checked that the skewsymmetric spin-orbit interaction vanishes for two particles of equal masses. Indeed, in this case, the wave function must either have a good symmetry under the exchange $1\leftrightarrow2$ if both particles are identical, or a good $C$-parity for a particle-antiparticle pair. In a two-body system, $\bm \Omega=0$ can be imposed by working in the center of mass frame.

If we considered an antiquark interacting with a quark or an antiquark, we would have found the same expression~(\ref{pot_1}), but with a different color factor, since the color operator would be different. These possible color operators are listed in eq.~(\ref{color:qq}), and denoted as ${\cal O}_{2,3;ijkl}$ respectively.

\subsection{Annihilation diagram}\label{anniqq}
In the particular case of the interaction between a quark and an antiquark of the same flavor, the contribution of a new diagram must be added, that is the annihilation diagram:
\begin{figure}[ht]
  \begin{center}
    \includegraphics*[width=3cm]{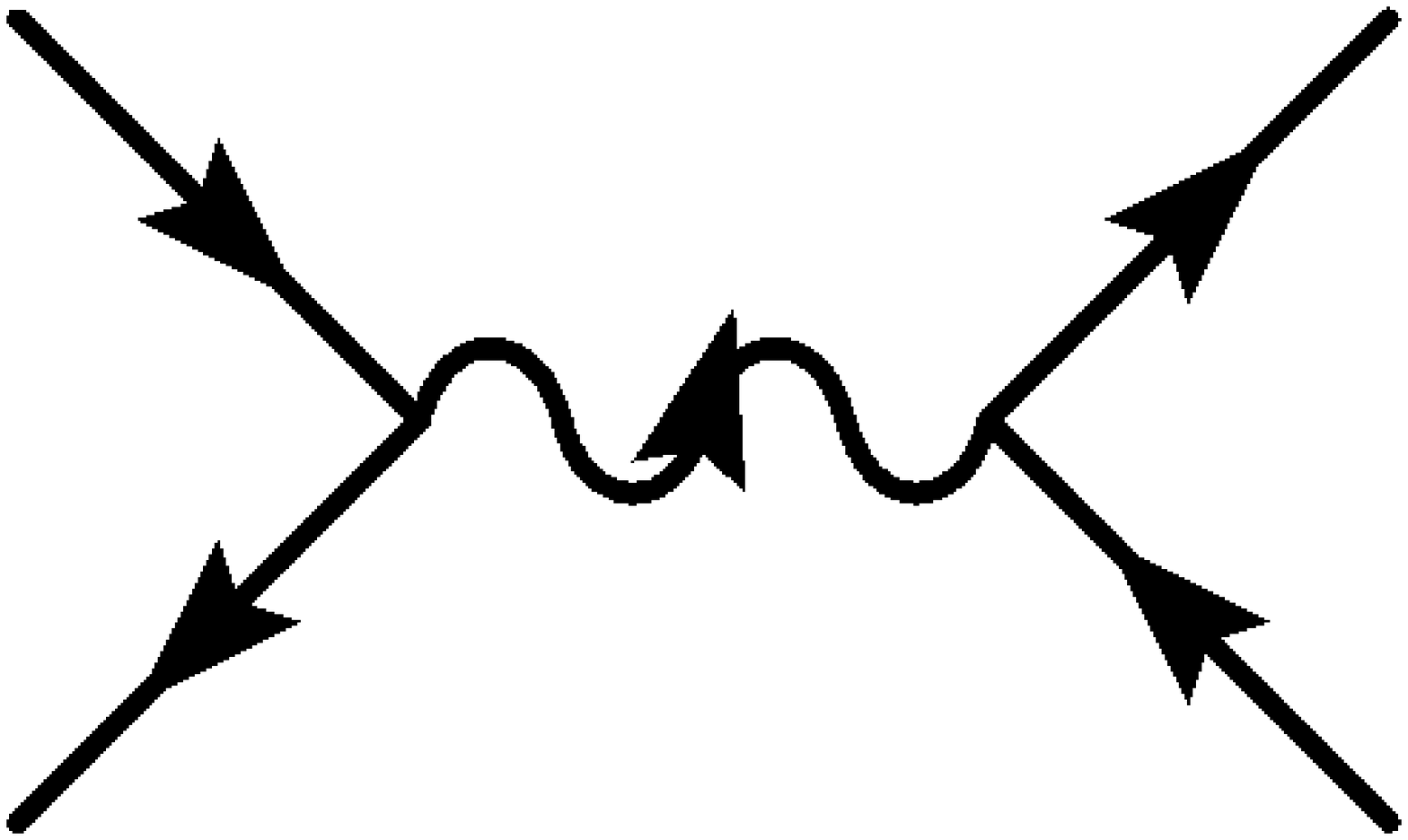}
  \end{center}
\end{figure}

\noindent The flavor of the outcoming pair is assumed to be identical to the incoming one. 

With the same indices as for the scattering diagram, we find
\begin{equation}\label{qqbstep1}
\begin{split}
    M_{fi}=&-i(2\pi)^4\delta^4(P'-P) (-g^2)  {\cal O}_{4,ijkl}\\
    &\times \left[\bar v_2\gamma^\mu\,  u_1\right] D_{\mu\nu}\left[\bar u'_1\gamma^\nu\,  v'_2\right],
   \end{split}
\end{equation}
with
\begin{equation}
    {\cal O}_{4;ijkl} = \frac{1}{4}(\lambda^a)_{ki}(\lambda_a)_{jl}.
\end{equation}
\par The independence of the matrix element~(\ref{qqbstep1}) in $\theta$ is readily checked. Consequently, the relevant part of the gluon propagator in the $s$-channel is given by
\begin{equation}\label{Ddef2}
    D_{\mu\nu}=\frac{\eta_{\mu\nu}}{q^2}\approx \frac{\eta_{\mu\nu}}{4m^2},
\end{equation}
because we can suppose that the momentum of the created gluon is very smaller that its rest energy. The wave functions~(\ref{uvdef}) can be taken at the lowest order, that is
\begin{eqnarray}
    u =\sqrt{2m}\ \begin{pmatrix}w\\
                            0\end{pmatrix},\quad
    v =\sqrt{2m}\  \begin{pmatrix}0\\
                                w\end{pmatrix}.
\end{eqnarray}
After a rearrangement of formula~(\ref{qqbstep1}) thanks to the relation
\begin{equation}    \sigma_{\alpha\beta}\sigma_{\delta\gamma}=-\frac{1}{2}\sigma_{\alpha\gamma}\sigma_{\delta\beta}+\frac{3}{2}\delta_{\alpha\gamma}\delta_{\delta\beta},
\end{equation}
one finds
\begin{equation}\label{qqbstep2}
M=\left(w'^\dagger_2\right)^{\rm c} w'^\dagger_1\   {\cal C}_4 \frac{g^2}{8m^2}\left[3+\bm\sigma_1\cdot\bm\sigma_2\right] \left(w_2\right)^{\rm c} w_1.
\end{equation}
${\cal O}_{4,ijkl}$ has been replaced by the corresponding factor ${\cal C}_4$, and we defined the charge conjugation as
\begin{equation}
    w^{\rm c}_\alpha=i\sigma_2 w^*_\alpha.
\end{equation}
After a Fourier transform, we get the annihilation potential
\begin{eqnarray}\label{pot_2}\nonumber
V(\bm r)&=& {\cal C}_4 \frac{2\pi\alpha_S}{m^2 }    \left[\frac{3}{4}+\bm S_1\cdot\bm S_2\right]\delta^3(\bm r),\\
    &=&{\cal C}_4 \frac{\pi\alpha_S}{m^2 }\bm S^2\delta^3(\bm r),
\end{eqnarray}
which has logically the form of a contact interaction. Moreover, it is a projector on a spin-$1$ state: The quark-antiquark pair must have the quantum numbers of a gluon in order to make the annihilation possible. The color factor ${\cal C}_4$ is indeed also a projector, on the octet state (see table~\ref{Tab:color:qq}).  Let us note that our potential~(\ref{pot_2}) is in agreement with the one of ref.~\cite{sem95}.

\subsection{Comparison with background perturbation theory}\label{bfapp}
The spin-dependent corrections coming from QCD are computed with the background perturbation theory in ref.~\cite{Simo00} for the $q\bar q$ pair in a meson. In our notations, the result reads
\begin{equation}\label{v_simo}
\begin{split}
    V^{SD}(\bm r)&=\left[\frac{\bm S_1\cdot\bm L_1}{2\mu^2_1}-\frac{\bm S_2\cdot\bm L_2}{2\mu^2_2}\right]\left[\frac{1}{r}\partial_r \epsilon(r)+\frac{2}{r}\partial_r V_1(r)\right]+\frac{\bm S_2\cdot\bm L_1-\bm S_1\cdot\bm L_2}{\mu_1\mu_2}\ \frac{1}{r}\partial_r V_2(r)
    \\&+\frac{\bm S_1\cdot\bm S_2}{3\mu_1\mu_2}V_4(\bm r)+\frac{1}{3\mu_1\mu_2}\left[3\frac{(\bm S_1\cdot\bm r)(\bm S_2\cdot\bm r)}{r^2}-\bm S_1\cdot\bm S_2\right]V_3(r),
\end{split}
\end{equation}
where $\epsilon(r)$ and $V_i(r)$ are rather complicated functions of the quark correlators. Fortunately, at the order $g^2$ they take simpler expressions, which are
\begin{eqnarray}\label{corr_def}
    V_1(r)=0, &V_2(r)=\epsilon(r)=-\frac{4}{3}\frac{\alpha_S}{r},\nonumber\\
    V_3(r)=\frac{4\alpha_S}{r^3}, &V_4(\bm r)=\frac{32}{3}\pi\alpha_S\delta^3(\bm r).
\end{eqnarray}
One can check that, by injecting the relations~(\ref{corr_def}) into the potential~(\ref{v_simo}), one formally obtains the spin-dependent part of the Fermi-Breit potential~(\ref{pot_1}) for a meson, that is for a color factor equal to ${\cal C}_2=-4/3$. The only difference is that, in the background field theory, the bare masses $m_\alpha$ are systematically replaced by $\mu_\alpha$, which are interpreted as dynamical, or constituent quark masses.

If one sees the potentials we are computing as perturbations of an effective Hamiltonian (typically a spinless Salpeter one with linear confining potential), the $\mu_\alpha$ can be defined as~\cite{bada02}
\begin{equation}\label{mudef}
    \mu_\alpha=\left\langle \sqrt{\bm p^2_\alpha+m^2_\alpha}\right\rangle,
\end{equation}
where the averages are computed with the unperturbed wave functions. The advantage of this approach is that the $\mu_\alpha$ are always nonzero; the development of the matrix elements in powers of $1/\mu_\alpha$ is thus relevant, even for light particles. Consequently, the equivalence between potentials~(\ref{pot_1}) and (\ref{v_simo}), suggest that in all our effective potentials, the quark masses should be seen as the dynamical ones, the $\mu_\alpha$ defined by eq.~(\ref{mudef}).
\par The same arguments also hold for gluons: Even if the exchanged gluons are seen as massless gauge particles, it is usually considered that the constituent gluons, in a glueball for example, acquire a mass because of the confining interaction. This point of view can also be checked with the background perturbation theory (see ref.~\cite{simo99}), and has already given good results in the description of glueballs~\cite{Brau04,Math}. Thus, in our case, the ``gluon mass" appearing in eq.~(\ref{epsdef1}) has to be understood as the constituent one.

\section{Gluon-Gluon interactions}\label{ggint}
\subsection{Scattering}\label{scatdiagg}

The computation of the effective potential coming from the scattering diagram between two gluons,
\begin{figure}[ht]
  \begin{center}
    \includegraphics*[width=3cm]{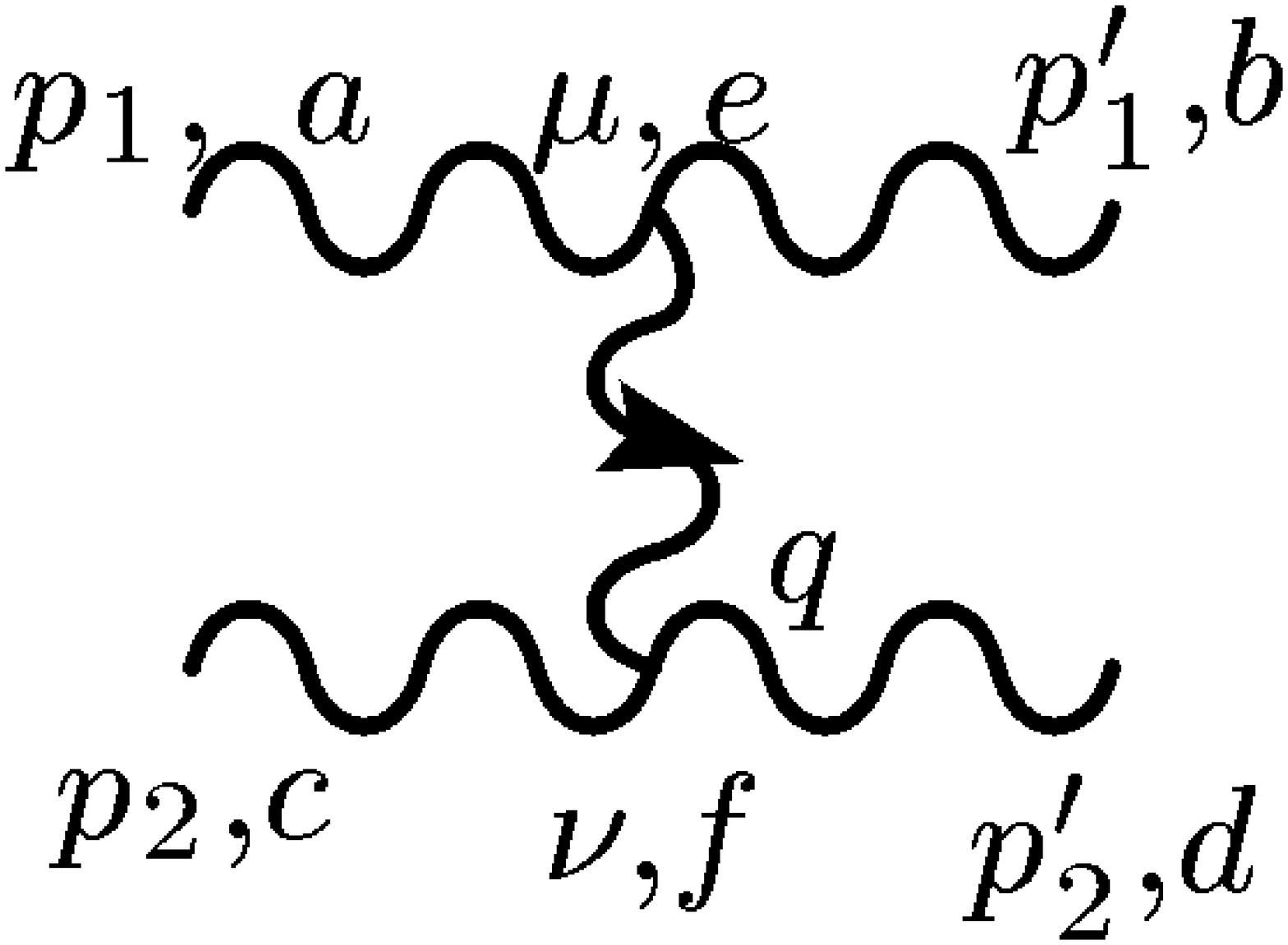}
  \end{center}
\end{figure}

\noindent is rather similar to the one of the scattering diagram between two quarks. The analog of formula~(\ref{qqstep1}) is now
\begin{equation}\label{Mscatdiag}
    M_{fi}=-i(2\pi)^4 \delta^4(P'-P) (-g^2)     {\cal O}_{5;abcd}\ G^\mu_1(q) D_{\mu\nu} G^\nu_2(-q),
\end{equation}
with the color operator
\begin{equation}
    {\cal O}_{5;abcd} =  -f_{\ ba}^ef_{edc},
\end{equation}
and the conserved current
\begin{eqnarray}\label{gdef}
G^\mu_\alpha(q)&=&(p_\alpha+p'_\alpha)^\mu\ (\epsilon_\alpha\cdot\epsilon^{*}_{\alpha})-\left[(q+p'_\alpha)\cdot\epsilon^{*}_{\alpha'}\right]\epsilon^{\mu}_\alpha\nonumber\\ &&+\left[(q-p'_\alpha)\cdot\epsilon_\alpha\right]\epsilon^{*\mu}_{\alpha}    .
\end{eqnarray}
With the aid of the definitions~(\ref{epsdef1}) and (\ref{spin1def}), and the property
\begin{equation}
(\bm a \cdot\bm \epsilon_\alpha)(\bm b\cdot\bm \epsilon^*_\alpha)-(\bm b \cdot\bm \epsilon_\alpha)(\bm a\cdot\bm \epsilon^*_\alpha) =i(\bm b\times\bm a)\cdot\bm S_\alpha(\bm\epsilon_\alpha\cdot\bm\epsilon^*_\alpha),
\end{equation}
 we can write the different components of $G^\mu_\alpha$ as
\begin{eqnarray}
    G^0_\alpha(\bm q)&=&-2p_\alpha^0\, \bm\epsilon_{\alpha}^*\left\{1-\frac{1}{2m^2c^2}\left[i(\bm q\times\bm p_\alpha)\cdot\bm S_\alpha +\bm q^2-(\bm S_\alpha\cdot  \bm q)^2\right]\right\}\bm \epsilon_\alpha\\  
\bm G_\alpha(\bm q)&=&2p_\alpha^0\left[-\frac{1}{2mc}(2\bm p_\alpha-\bm q)\cdot\bm\epsilon_{\alpha}^{*}\, \bm\epsilon_\alpha+\frac{\bm q\cdot\bm\epsilon^*_{\alpha}}{mc}\ \bm\epsilon_\alpha - \frac{\bm q\cdot\bm\epsilon_\alpha}{mc}\, \bm\epsilon^*_{\alpha}\right].
\end{eqnarray}
As the gluon propagator is given by relations~(\ref{propadef}), the matrix element~(\ref{Mscatdiag}) becomes
\begin{equation}\label{Mscatdia2}
\begin{split}
    M_{fi}=-i(2\pi)^4 \delta^4(P'-P)\  {\cal C}_{5}\,  \frac{g^2}{\bm q^2} \left[G^0_1G^0_2-\bm G_1\cdot\bm G_2+\frac{(\bm q\cdot\bm G_1)(\bm q\cdot\bm G_2)}{\bm q^2}  \right],
\end{split}
\end{equation}
where we used the obvious notation $G_1=G_1(\bm q)$ and $G_2=G_2(-\bm q)$. By simplification of eq.~(\ref{Mscatdia2}), and with formula~(\ref{spin1prop}), one can arrive at the analog of eq.~(\ref{qqstep2}), that is
\begin{equation}
M=(4p^0_1p^0_2)\,  \bm\epsilon_{1}^*\bm\epsilon_{2}^*\, \left[g^2\, {\cal C}_5\, U(\bm q,\bm p_1,\bm p_2)\right]\bm\epsilon_1\bm\epsilon_2,
\end{equation}
with
\begin{eqnarray}\label{inter1}
        U(\bm q,\bm p_1,\bm p_2)& =&\frac{1}{\bm q^2}-\frac{1}{m^2}-\frac{i\bm S_1\cdot(\bm q\times\bm p_1)}{2m^2\bm q^2}+\frac{i\bm S_2\cdot(\bm q\times\bm p_2)}{2m^2\bm q^2}+\frac{(\bm p_1\cdot\bm q)( \bm p_2\cdot\bm q)}{m^2\bm q^4}-\frac{\bm p_1\cdot\bm p_2}{m^2\bm q^2} \nonumber\\
        &&+\frac{i\bm S_1\cdot(\bm q\times \bm p_2)}{m^2\bm q^2}-\frac{i\bm S_2\cdot(\bm q\times \bm p_1)}{m^2\bm q^2} +\frac{(\bm S_1\cdot\bm q)(\bm S_2\cdot\bm q)}{m^2\bm q^2}-\frac{\bm S_1\cdot\bm S_2}{m^2}\nonumber\\
&&+\frac{(\bm S_1\cdot\bm q)^2+(\bm S_2\cdot\bm q)^2}{2m^2\bm q^2}  .
\end{eqnarray}
The constituent gluons are assumed to have the same effective mass: This is clear because of the symmetry of the problem.
The effective potential in position space is thus
\begin{eqnarray}\label{pot_3}
    V(\bm r)&=&\, {\cal C}_5\, \alpha_S\left\{\frac{1}{r}-\frac{4\pi}{3m^2}\delta^3(\bm r)-\frac{\bm L_1\cdot\bm S_1}{2m^2r^3}+
    \frac{\bm L_2\cdot\bm S_2}{2m^2r^3}- \frac{1}{2m^2r}\left[\bm p_1\cdot\bm p_2 +\frac{(\bm p_1\cdot\bm r)(\bm p_2\cdot\bm r)}{r^2}\right]\right.\nonumber\\
    &&-\frac{1}{m^2r^3}\left[ \bm L_1\cdot\bm S_2 - \bm L_2\cdot\bm S_1\right]+ \frac{1}{m^2}\left[\frac{\bm S_1\cdot\bm S_2}{r^3}-3\frac{(\bm S_1\cdot\bm r)(\bm S_2\cdot\bm r)}{ r^5}\right]-\frac{8\pi}{3m^2}\bm S_1\cdot\bm S_2\delta^3(\bm r)\nonumber\\
    &&\left.+\frac{2}{m^2r^3}-\frac{3}{2m^2r^5}\left[(\bm S_1\cdot\bm r)^2+(\bm S_2\cdot\bm r)^2\right]\right\}.
\end{eqnarray}
This potential is very similar to~(\ref{pot_1}), except in two terms. The first one is the contact interaction, which is stronger than in the quark-quark case, and the second one is, at first sight, a new interaction term, written in the last line of eq.~(\ref{pot_3}). The existence of this term has already been pointed out in ref.~\cite{bar}, but here we go further than this reference because we completely remove the $\bm \epsilon_\alpha$ from our expression. Actually, these terms are a part of the tensor force, because if we rewrite the potential (\ref{pot_3}) with the new variables defined by (\ref{change}), we obtain
\begin{eqnarray}\label{pot_3bis}
    V(\bm r)&=&{\cal C}_5 \alpha_S\left\{\frac{1}{r}+\frac{4\pi}{3m^2}\left(3-\bm S^2\right)\delta^3(\bm r)-\frac{1}{2m^2}\left[\frac{\bm p_1\cdot\bm p_2}{r}+\frac{(\bm p_1\cdot\bm r)(\bm p_2\cdot\bm r)}{r^3}\right]\right.\nonumber\\
    &&-\frac{3}{2m^2}\frac{\bm L\cdot\bm S }{r^3}+\frac{1}{4m^2}\frac{\bm \delta\cdot\bm \Omega}{r^3}+\left.\frac{1}{2m^2}\left[\frac{\bm S^2}{r^3}-3\frac{(\bm S\cdot\bm r)^2}{r^5}\right]\right\}.
\end{eqnarray}

\subsection{Annihilation}\label{annigg}

The annihilation of two gluons, represented by the diagram
\begin{figure}[ht]
  \begin{center}
    \includegraphics*[width=3cm]{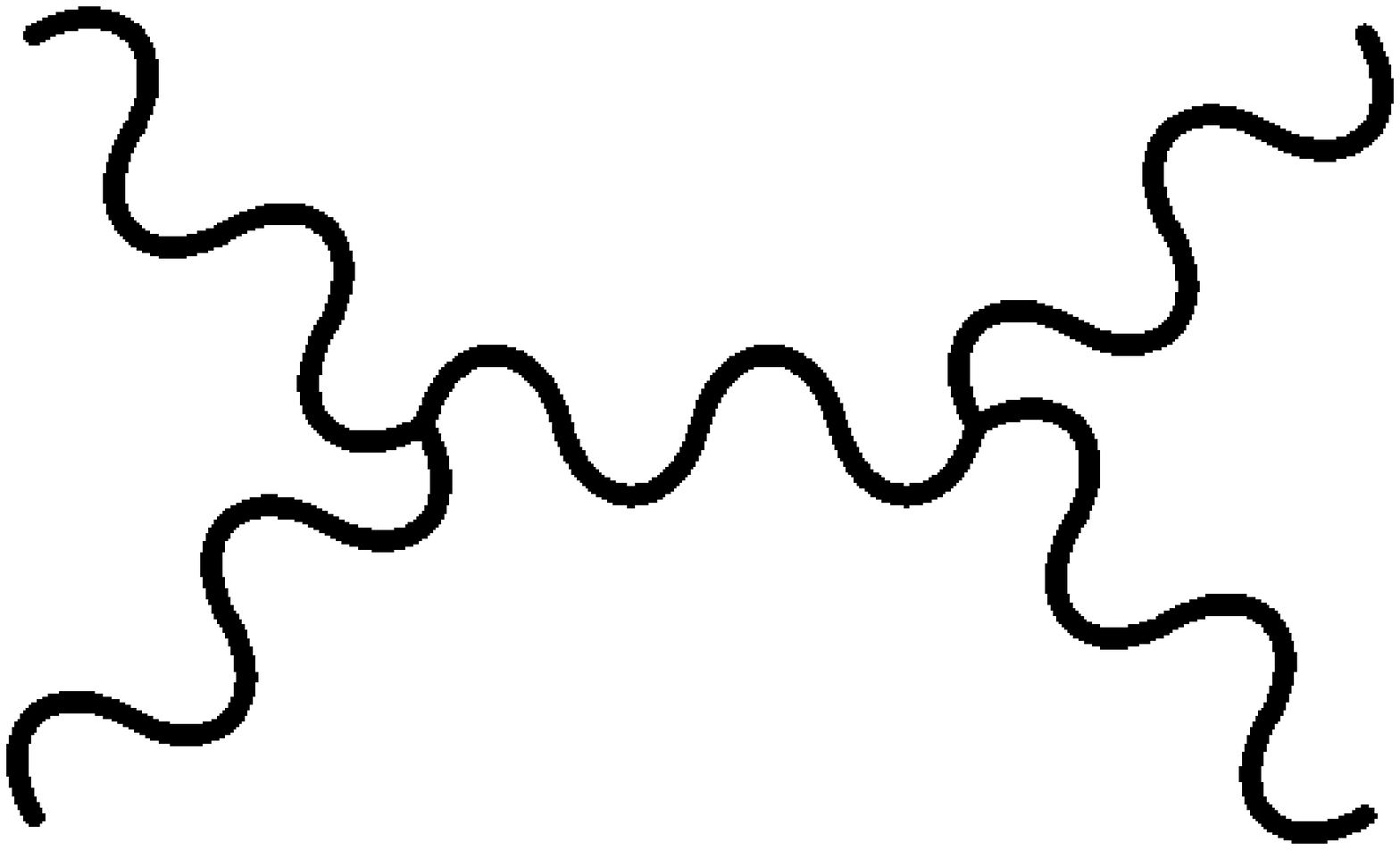}
  \end{center}
\end{figure}

\noindent is a special case because its contribution is always zero. First of all, the color operator associated to this diagram is
\begin{equation}
    {\cal O}_{6;abcd} =  f_{\ ca}^ef_{ebd}.
\end{equation}
It is shown in sec.~\ref{colfactc} that the corresponding color factor is non zero only if the gluon pair is in an antisymmetric color octet. But even in this case, a calculation similar to the one performed in sec.~\ref{anniqq} shows that the matrix element is zero at the order $c^{-2}$. Indeed, as the gluon propagator in the $s$-channel can be taken of the form~(\ref{Ddef2}), the wave functions and momenta will simply be given by their lowest order approximation
\begin{equation}\label{epslow}
\epsilon=(0,\bm\epsilon), \quad p=(m,\bm 0),
\end{equation}
 respectively. This approximation causes the matrix element to vanish in every case, as it is mentioned in ref.~\cite{hou}.

\subsection{Seagull}\label{conta}

The seagull diagram possesses a unique vertex: The four-gluon one.

\begin{figure}[ht]
  \begin{center}
    \includegraphics*[width=2cm]{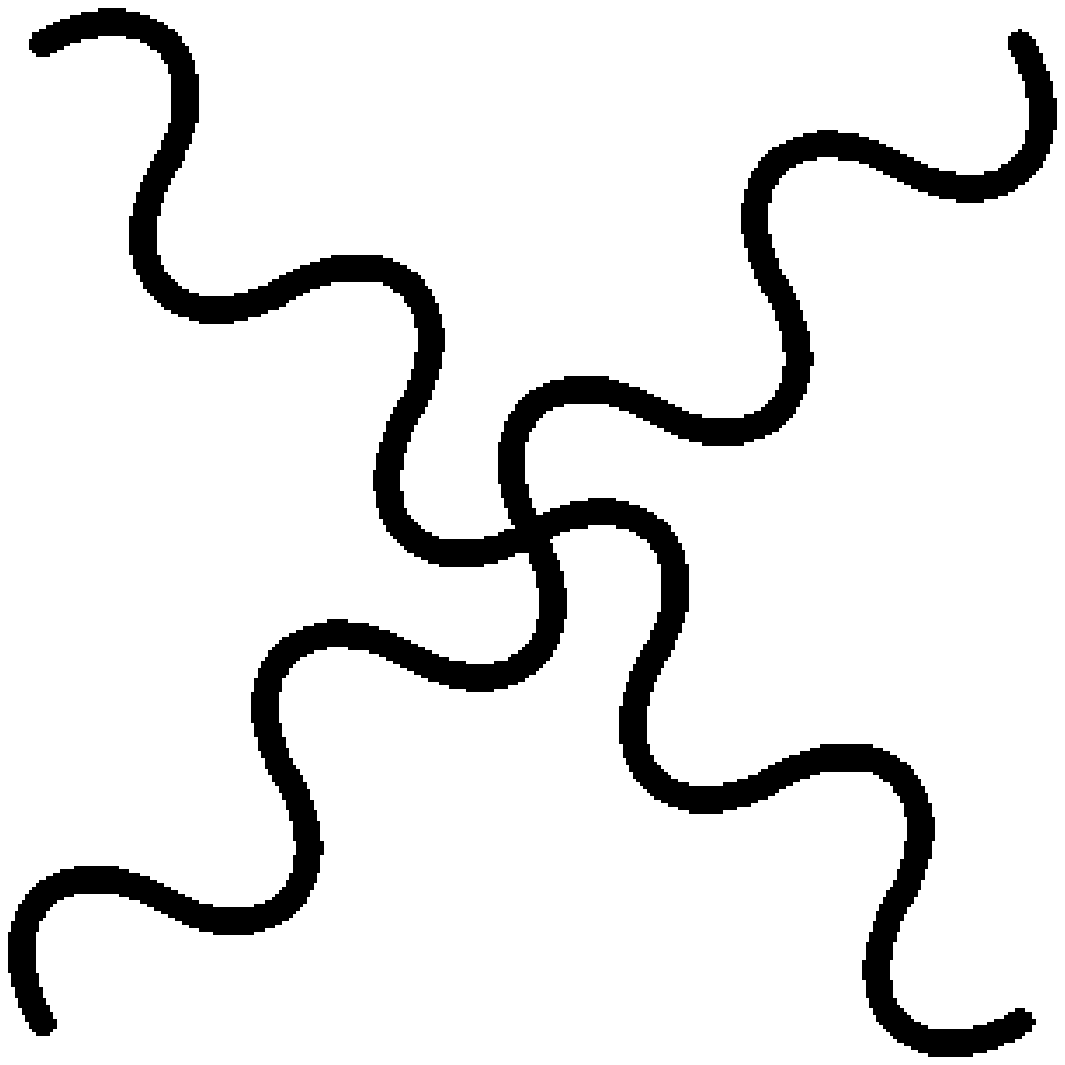}
  \end{center}
\end{figure}

\noindent Applying the corresponding Feynman rules together with the Jacobi identity
\begin{equation}
    f^e_{ab}f_{ecd}+    f^e_{ac}f_{edb}+    f^e_{ad}f_{ebc}=0,
\end{equation}
we obtain the following matrix element
\begin{eqnarray}\label{sgstep1}
    M_{fi}&=&-i(2\pi)^4\delta^4(P'-P) \frac{-g^2}{2} \left\{{\cal O}_{6;abcd}(2\epsilon_{1}^*\cdot\epsilon_{1}\ \epsilon_{2}^*\cdot\epsilon_{2}-\epsilon_{1}^*\cdot\epsilon_{2}\ \epsilon_{2}^*\cdot\epsilon_{1} - \epsilon_{1}^*\cdot\epsilon_{2}^*\ \epsilon_{1}\cdot\epsilon_{2})\right.\nonumber\\
    &&+\left.{\cal O}_{5;abcd}(
    \epsilon_{1}^*\cdot\epsilon_{1}\ \epsilon_{2}^*\cdot\epsilon_{2}-\epsilon_{2}^*\cdot\epsilon_{1}\ \epsilon_{1}^*\cdot\epsilon_{2})\right\}.
\end{eqnarray}
An additional $1/2$ factor is present in eq.~(\ref{sgstep1}). This is due to the degeneracy of the different diagrams. Indeed, if we attribute a weight $1$ to the gluon-gluon scattering diagram, we have to attribute a weight $1/2$ to the annihilation and to the seagull diagrams \cite{bar}.
\par The effective potential corresponding to the seagull diagram should clearly be a contact-like interaction. It is thus sufficient to inject the lowest-order expressions~(\ref{epslow}) to simplify eq.~(\ref{sgstep1}). The following relations have also to be used
\begin{subequations}\label{propsg}
\begin{eqnarray}
(\bm\epsilon_{1}^*\cdot\bm\epsilon_{2}) (\bm\epsilon_{2}^*\cdot\bm\epsilon_{1}) &=&\bm\epsilon^*_{2}\bm\epsilon^*_{1}\, (\bm S_1\cdot\bm S_2)\, \bm\epsilon_2\bm\epsilon_1 + (\bm\epsilon_{1}^*\cdot\bm\epsilon_{2}^*) (\bm\epsilon_{1}\cdot\bm\epsilon_{2}),\\
    {\rm Tr}\left[(\bm\epsilon^*_\alpha \bm S_\alpha).(\bm S_\alpha\bm\epsilon_\alpha)\right] &=& \bm\epsilon^*_\alpha\bm S^2_\alpha\bm\epsilon_\alpha.
\end{eqnarray}
Moreover,
\begin{equation}
(\bm\epsilon^*_1 \bm S)_b(\bm\epsilon_1\bm S)_{a}(\bm\epsilon^*_2 \bm S)^b(\bm\epsilon_2\bm S)^{a}=
    \bm\epsilon^*_1\bm\epsilon_2^* (\bm S_1\cdot\bm S_2)^2 \bm\epsilon_1\bm\epsilon_2\approx0,
\end{equation}
\end{subequations}
since such terms are higher order relativistic corrections.
The effective potential in momentum space reads
\begin{equation}
    U(\bm q) =\frac{g^2}{8m^2}\left\{{\cal C}_6 \left[\bm S_1\cdot\bm S_2-4\right]
+   {\cal C}_5 \left[\bm S_1\cdot\bm S_2-2\right]\right\}.
\end{equation}
Introducing the total spin of the system, we get in position space
\begin{equation}\label{pot_sg}
    V(\bm r) =\frac{\pi\alpha_S}{2m^2}\left\{{\cal C}_6 \left[\frac{1}{2}\bm S^2-6\right]+
    {\cal C}_5 \left[\frac{1}{2}\bm S^2-4\right]\right\}\delta^3(\bm r).
\end{equation}
The complete potential to use to describe the interactions between two gluons is thus the sum of potentials (\ref{pot_3bis}) and (\ref{pot_sg}).

\section{(Anti)quark - Gluon interactions}\label{qgint}

The understanding of hybrid mesons currently deserves much interest in experimental as well as in theoretical physics. From a quark model point of view, hybrid mesons can be seen as a quark-antiquark-constituent gluon bound state. In the framework of a potential model, the dominant part of the interaction is a confining term which is typically given by $a|\bm x_q-\bm x_g|+a|\bm x_{\bar q}-\bm x_g|$~\cite{horn}. Then, short-range interactions should be added in order to build a spin-dependent model. The quark-antiquark potential has been computed in Sec.~\ref{qqint}; in this case, the $q\bar q$ pair is in a color octet and couples to the constituent gluon in order for the total system to be in a color singlet. But, the short-range potentials between the constituent gluon and the quark or the antiquark have, up to our knowledge, never been computed. Consequently, the knowledge of (anti)quark-gluon interactions is a missing element in the building of accurate potential models of hybrid mesons. That is why we compute these interactions in the present section, applying the same procedure as the one used up to now.    

\subsection{Scattering diagram}

There are three diagrams to consider when one deals with quark-gluon interactions \cite[p. 196]{dere}. We begin with the following scattering diagram,
\begin{figure}[ht]
  \begin{center}
    \includegraphics*[width=2.5cm]{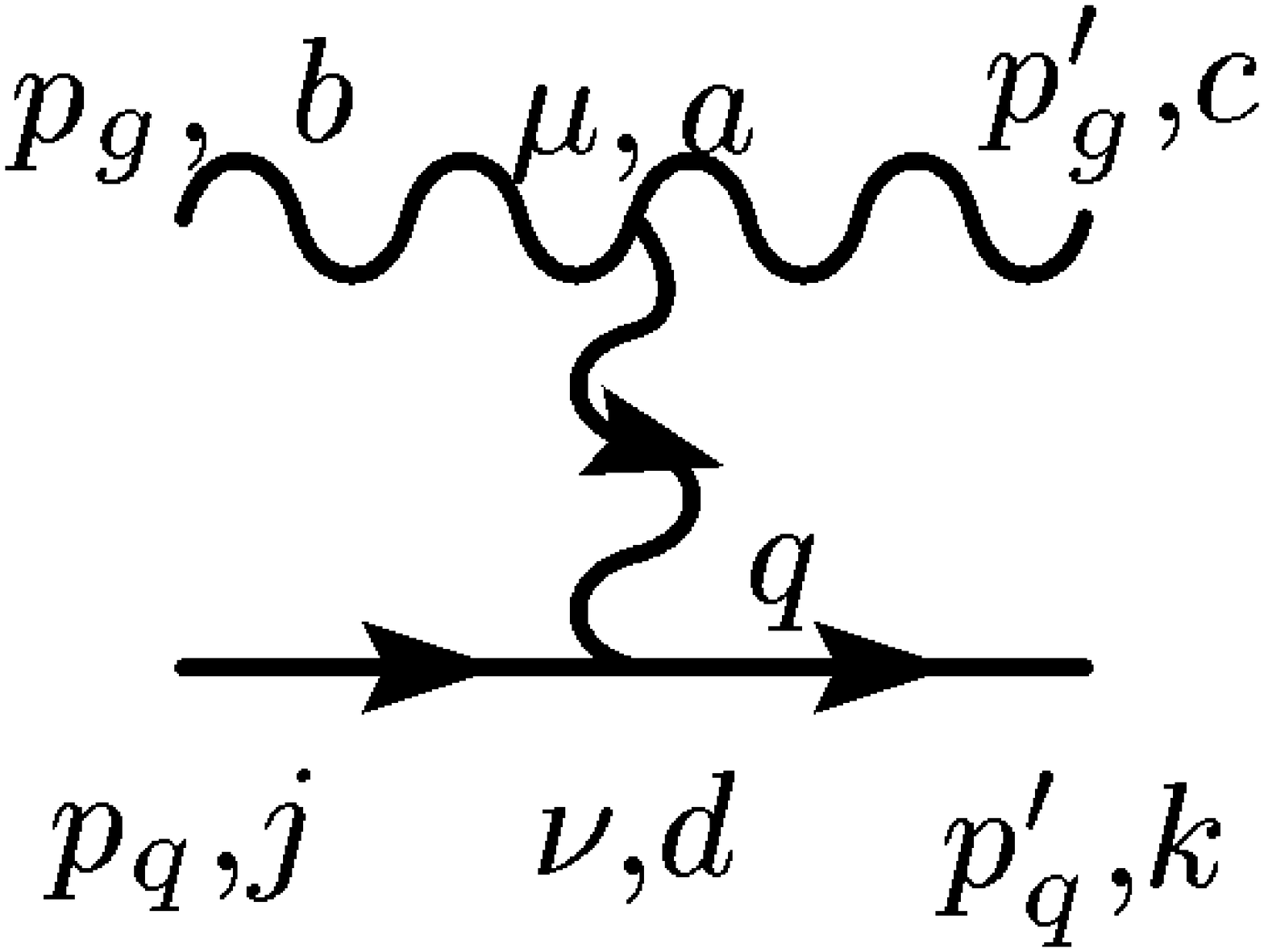}
  \end{center}
\end{figure}

\noindent In the following, the indices $1,2$ will be replaced by $g,q$ in order to clearly distinguish between the symbols related to the gluon and to the quark respectively.

Applying the Feynman rules, we can write the corresponding matrix element in form \eqref{mif0},
\begin{equation}
    M_{fi}=-i(2\pi)^4 \delta^4(P'-P) g^2    {\cal O}_{7;bcjk}\ G^\mu_g(q) D_{\mu\nu} J^\nu_q(-q),
\end{equation}
where the color factor is
\begin{equation}
    {\cal O}_{7;bcjk} = \left(\frac{\lambda^a}{2}\right)_{kj}(-if_{acb}),
\end{equation}
and where $J_q^\nu(-q)$ and $G^\mu_g(q)$ were already computed in eqs.~\eqref{Fmu} and \eqref{gdef}. Using the propagator (\ref{propadef}), the matrix element can rewritten as
\begin{eqnarray}\label{qgmfi}
    M_{fi}&=&-i(2\pi)^4 \delta^4(P'-P)\  {\cal C}_{7}\,  \frac{g^2}{\bm q^2} \left[-G^0_gJ^0_q+\bm G_g\cdot\bm J_q-\frac{(\bm q\cdot\bm G_g)(\bm q\cdot\bm J_q)}{\bm q^2}  \right].
\end{eqnarray}
After the same rearrangements as those which were done in secs. \ref{qqint} and \ref{ggint}, eq.~(\ref{qgmfi}) becomes
\begin{equation}
    M=(4p^0_gm_q)\,\bm\epsilon^*w'^\dag\, \left[g^2\, {\cal C}_7\, U(\bm q,\bm p_g,\bm  p_q)\right]\bm\epsilon w,
\end{equation}
with
\begin{eqnarray}\label{pot_40}
        U(\bm q,\bm p_g,\bm p_q) &=&\frac{1}{\bm q^2}-\left(\frac{1}{2m^2_g}+\frac{1}{4m_q^2}\right)-\frac{i\bm S_g\cdot(\bm q\times\bm p_g)}{2m_g^2\bm q^2}+\frac{i\bm S_q\cdot(\bm q\times\bm p_q)}{2m_q^2\bm q^2}\nonumber\\
        &&+\frac{(\bm p_g\cdot\bm q)(\bm p_q\cdot\bm q)}{m_gm_q\bm q^4}-\frac{\bm p_g\cdot\bm p_q}{m_gm_q\bm q^2}+\frac{i\bm S_g\cdot(\bm q\times\bm p_q)}{m_gm_q\bm q^2}-\frac{i\bm S_q\cdot(\bm q\times\bm p_g )}{m_qm_g\bm q^2}\nonumber\\
        &&+\frac{(\bm S_g\cdot\bm q)(\bm S_q\cdot\bm q)}{m_gm_q\bm q^2}-\frac{\bm S_q\cdot\bm S_g}{m_gm_q}+\frac{(\bm S_g\cdot\bm q)^2}{2m_g^2\bm q^2}+\frac{(\bm S_q\cdot\bm q)^2}{2m_q^2\bm q^2}.
\end{eqnarray}
One can check that the quark part of potential~(\ref{pot_40}) is equal to the one of eq.~(\ref{eq22}), and that the gluon part is the same as in eq.~(\ref{inter1}), which is rather coherent. The last term of this expression is actually not explicitly written in the quark-quark case. It is indeed eliminated thanks to the properties of the Pauli matrices through the relation $$(\bm{ S_q \cdot q})^2/\bm q^2=1/4;$$ consequently, it is only a constant term, which will be part of the contact interaction after a Fourier transform. In the gluon-gluon scattering, the gluonic term in $(\bm{ S_g \cdot q})^2$ was naturally included in the tensor force. In the quark-gluon case, the masses $m_g,m_q$ are not equal, and this term has thus to be treated separately. 

In position space, the effective potential is given by
\begin{eqnarray}
        V(\bm r) &=&\ {\cal C}_7\alpha_S\left\{\frac{1}{r}-\left(\frac{2\pi}{3m^2_g}+\frac{\pi}{2m_q^2}\right)\delta^3(\bm r)- \frac{\bm L_g\cdot\bm S_g}{2m_g^2r^3}+\frac{\bm L_q\cdot\bm S_q}{2m_q^2r^3}- \frac{8\pi}{3m_gm_q}\bm S_g\cdot\bm S_q\delta^3(\bm r)\right.\nonumber\\
        &&-\frac{1}{2m_gm_q}\left[\frac{\bm p_g\cdot\bm p_q}{r} + \frac{(\bm p_g\cdot\bm r)(\bm p_q\cdot\bm r)}{r^3}\right]-\frac{1}{m_gm_qr^3}\left[\bm L_g\cdot\bm S_q-\bm L_q\cdot\bm S_g\right]\nonumber\\
        && +\frac{1}{m_gm_q}\left[\frac{\bm S_g\cdot\bm S_q}{r^3}-3\frac{(\bm S_g\cdot\bm r)(\bm S_q\cdot\bm r)}{r^5}\right]\left.
        +\frac{1}{m_g^2r^3}-\frac{3}{2m_g^2}\frac{(\bm S_g\bm r)^2}{r^5}\right\}.
\end{eqnarray}
Rewriting it with the appropriate variables we have
\begin{eqnarray}\label{pot_4}
        V(\bm r) &=&\, {\cal C}_7\alpha_S\left\{\frac{1}{r}-\frac{1}{2m_gm_q}\left[\frac{\bm p_g\cdot\bm p_q}{r} + \frac{(\bm p_g\cdot\bm r)(\bm p_q\cdot\bm r)}{r^3}\right]\right.\nonumber\\
        &&+\pi\left(\frac{11}{3m_gm_q}-\frac{2}{3m^2_g}-\frac{1}{2m^2_q}-\frac{4}{3m_gm_q}\bm S^2\right)\delta^3(\bm r)-\left(\frac{m^2_q+m^2_g+4m_gm_q}{4m_g^2m_q^2}\right)\frac{\bm L\cdot\bm S }{r^3}\nonumber\\
        &&+\frac{1}{4m_gm_q}\frac{\bm \delta\cdot\bm \Omega}{r^3}+\frac{(m_q-m_g)}{4m_qm_g(m_q+m_g)}\frac{\bm S\cdot\bm \Omega}{r^3}
    +\left(\frac{m_g^2-m_q^2}{4m_q^2m_g^2}\right)\frac{\bm L\cdot\bm \delta}{r^3}\nonumber\\
    &&\left.+\frac{1}{2m_gm_q}\left[\frac{\bm S^2}{r^3}-3\frac{(\bm S\cdot\bm r)^2}{r^5}\right]+\frac{m_q-m_g}{m^2_gm_q}{\cal Q}(\bm r)\right\},
\end{eqnarray}
with
\begin{equation}\label{q1}
{\cal Q}(\bm r)=    \left[\frac{1}{r^3}-\frac{3(\bm S_g\cdot\bm r)^2}{2r^5}\right].
\end{equation}
The physical meaning of this term can be clarified by using the following relation involving the spin-$1$ matrices \cite[p. 55]{varsha}
\begin{equation}
    \bm S_{gi}\bm S_{gk}=\frac{2}{3}\delta_{ik}+\frac{i}{2}\epsilon_{ik}^{\ \ l}\bm S_{gl}+Q_{ik}.
\end{equation}
$Q_{ik}$ are the polarisation matrices, defining the quadrupole tensor. Consequently, eq.~(\ref{q1}) can be rewritten as
\begin{equation}
    {\cal Q}(\bm r)=-\frac{3}{2}\frac{\bm r Q\bm r}{r^5}.
\end{equation}
A quadrupolar interaction is thus present in the potential. Such a term is logically absent in symmetric configurations, i.e. quark-quark and gluon-gluon systems, but appears in this asymmetric quark-gluon system.

\subsection{``Compton" diagrams}
The two remaining diagrams are analog to those of the Compton scattering in QED. We begin by the following annihilation diagram,
\begin{figure}[ht]
  \begin{center}
    \includegraphics*[width=3.0cm]{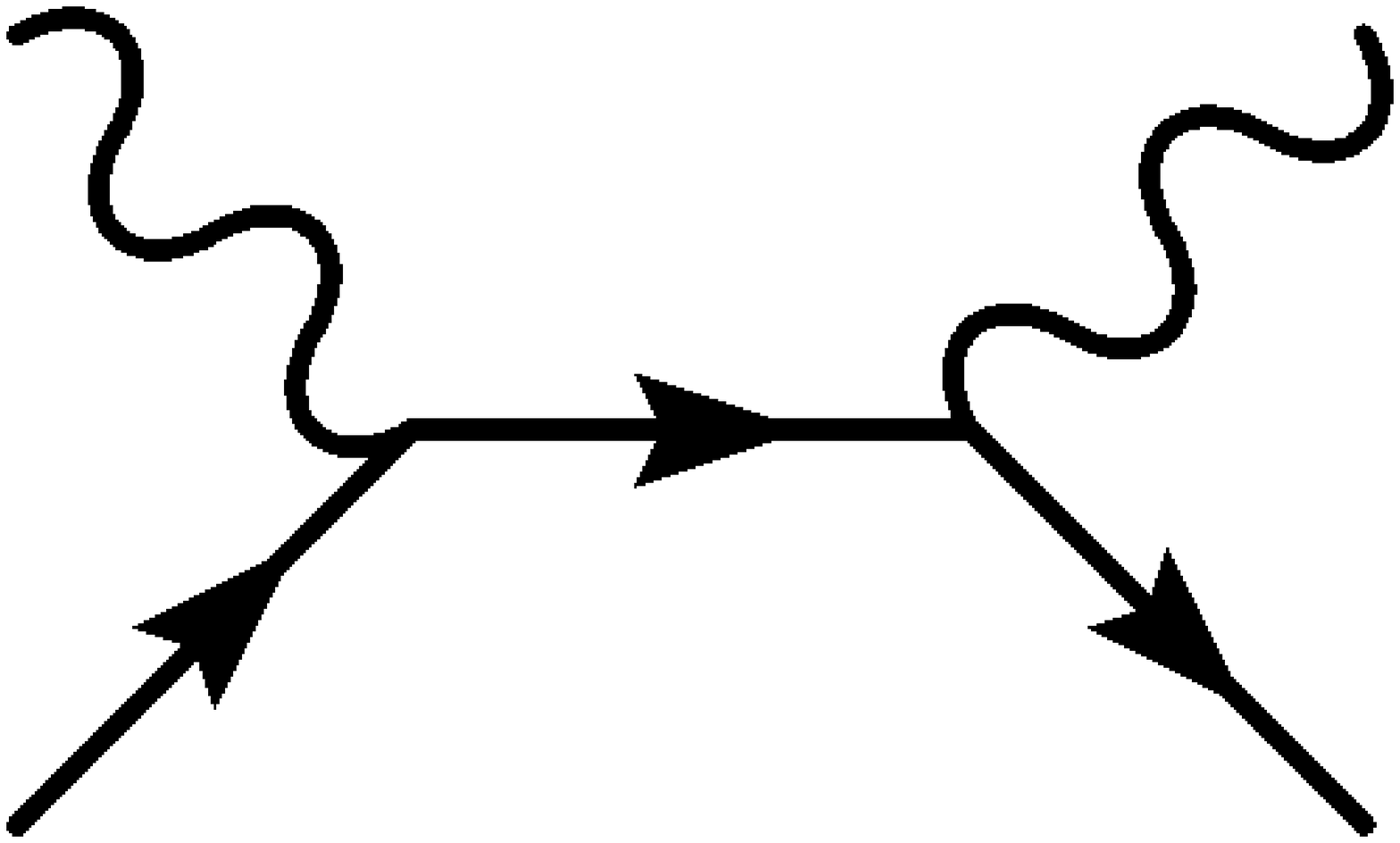}
  \end{center}
\end{figure}

\noindent whose matrix element reads
\begin{equation}\label{qgstep1}
    M_{fi}=-i(2\pi)^4 \delta^4(P'-P) g^2    {\cal O}_{8;bcjk}\ \bar{u}\, \epsilon\!\!\!/^*\frac{(q\!\!\!/+\tilde{m})}{q^2-\tilde{m}^2}\epsilon\!\!\!/\ u\, ,
\end{equation}
where the color factor is
\begin{equation}
    {\cal O}_{8;bcjk} = \frac{1}{4} (\lambda^c\lambda^b)_{kj}.
\end{equation}
The quark propagator reads
\begin{equation}
\frac{(q\!\!\!/+\tilde{m})}{q^2-\tilde{m}^2} \approx\frac{(m_g+m_q)\gamma^0+\tilde{m}\mathun}{(m_g+m_q)^2-\tilde{m}^2},
\end{equation}
in the approximation $q=(m_g+m_q, \bm0)$. $\tilde{m}$ is the mass of the propagating quark, which we choose to be equal to $m_q$. Using the lowest order approximations for the wave functions and the momenta, and thanks to formula
\begin{equation}
    (\bm \epsilon'\cdot\bm\sigma)(\bm \epsilon\cdot \bm \sigma)=\bm \epsilon'\cdot\bm \epsilon-\bm \epsilon'(\bm S\cdot\bm\sigma)\bm \epsilon,
\end{equation}
we find
\begin{equation}
M=4p^0_g m_q\,  w'^\dagger\, \bm\epsilon^*\left[{\cal C}_8 g^2 \frac{1-2\bm S_g\cdot\bm S_q}{2m_g(m_g+2m_q)}\right] w\bm\epsilon.
\end{equation}

The effective potential,
\begin{equation}\label{qgpot2}
    V(\bm r) =  \frac{2\pi{\cal C}_8 \alpha_S}{m_g(m_g+2m_q)}\left[\frac{15}{4}-\bm S^2\right]\delta^3(\bm r),
\end{equation}
is a projector on a spin $1/2$ state because the quark-gluon system must have the quantum numbers of a quark in order to form a single quark. If one now considers an antiquark instead of a quark, the only modification in the potential (\ref{qgpot2}) will be the color operator, given by
\begin{equation}
    {\cal O}_{9;bcjk} = \frac{1}{4} (\lambda^b\lambda^c)_{jk}.
\end{equation}

\par The second Compton-like diagram is
\pagebreak
\begin{figure}[ht]
  \begin{center}
    \includegraphics*[width=6.0cm]{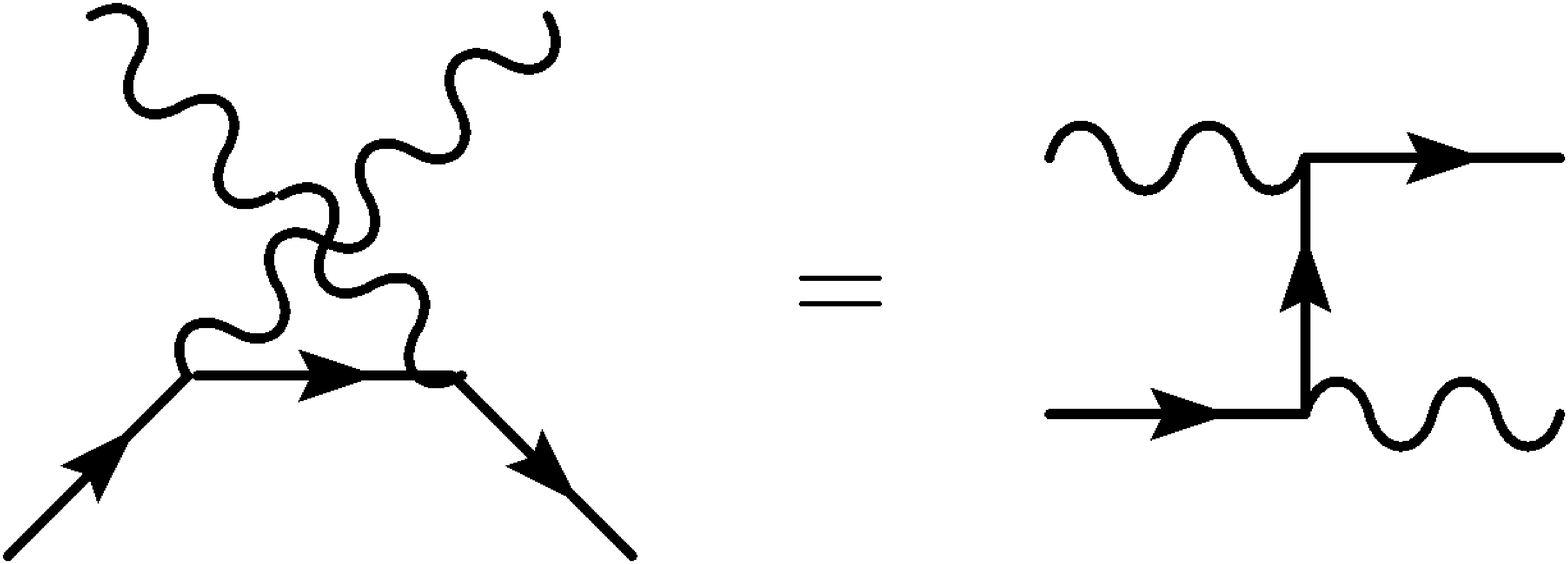}
  \end{center}
\end{figure}

\noindent This diagram is no longer an annihilation one. It can indeed be seen as an exchange of a(n) (anti)quark between a (anti)quark-gluon pair. We will thus work in Breit's frame, as in the case of the usual scattering diagrams: $q=(0,\bm q)$.  The exchanged quark has a constituent mass $m_q$, which is always nonzero in our formalism. Consequently, we expect to find an expression in which a Yukawa potential appears, whose contribution will in every case be less important than the one of the quark-gluon scattering diagram since the exchanged gluons are massless. This leads us to assume that only the lowest order expression of the effective potential will be relevant. One obtains easily for this potential
\begin{equation}\label{qgpot3}
    V(\bm r) =  {\cal C}_{10} \alpha_S\frac{m_q}{2m_g}\left[\bm S^2-\frac{7}{4}\right]\frac{{\rm e}^{-m_q r}}{r},
\end{equation}
${\cal C}_{10}$ corresponding to the color operator
\begin{equation}
    {\cal O}_{10;bcjk} = \frac{1}{4} (\lambda^b\lambda^c)_{kj}.
\end{equation}

Let us remark that if one deals with a quark instead of an antiquark in this diagram, the only change in potential (\ref{qgpot3}) will be again the color operator, given by
\begin{equation}
    {\cal O}_{11;bcjk} =\frac{1}{4} (\lambda^c\lambda^b)_{jk} .
\end{equation}

\section{Gluons: massive or not massive}\label{applic}

In the computation of all the potentials that we presented up to now, the constituent gluons were considered as massive particles because of confinement, but the exchanged gluons remained massless gauge particles. The advantage of this assumption, already used in ref.~\cite{bar}, is to treat the diagrams involving constituent (anti)quarks and/or gluons on an equal footing. Moreover, it allows to compute the potentials with the standard Feynman rules corresponding to the QCD Lagrangian. 

However, in some other approaches related to potential models of glueballs, it is argued that the exchanged gluons have to be massive as well in order for all the gluons to have an equal status \cite{corn,hou}. In these references, consequently, a gluonic mass term has to be added to the QCD Lagrangian. For completeness, we discuss here the consequences of such a term. Actually, the Feynman rules will not be modified at the level of the gluon propagator: the Coulomb form in position space, denoted hereafter as $U(r)=1/r$, will turn into the propagator of a massive vector particle, typically $\tilde U(r)=e^{-mr}/r$. Thus, at the first order in the relativistic corrections, only the scattering diagrams will be affected by this modification. The corresponding scattering potentials, denoted here as $V^S$, and given by eqs.~(\ref{pot_1bis}), (\ref{pot_3bis}), and (\ref{pot_4}), can be rewritten as
\begin{eqnarray}\label{sqqpot}
    V^S_{qq}&=&{\cal C}_1 \alpha_S\left\{U-\left(\frac{3-2\bm S^2}{6m_1m_2}-\frac{1}{8m^2_1}-\frac{1}{8m^2_2}\right)\bm\nabla^2U-\frac{1}{2m_1m_2}\left[\bm p_1U\bm p_2-(\bm p_1\cdot\bm r)\frac{U'}{r}(\bm r\cdot\bm p_2)\right]\right.\nonumber\\
    &&+\left(\frac{m^2_2+m^2_1+4m_1m_2}{4m_1^2m_2^2}\right)\bm L\cdot\bm S \ \frac{U'}{r}-\frac{1}{4m_1m_2}\bm \delta\cdot\bm \Omega\ \frac{U'}{r}-
    \left(\frac{m_1^2-m_2^2}{4m_2^2m_1^2}\right)\bm L\cdot\bm \delta\ \frac{U'}{r}\nonumber\\
    &&-\frac{(m_2-m_1)}{4(m_1+m_2)m_1m_2}\bm S\cdot\bm \Omega\ \frac{U'}{r}+\left.\frac{1}{6m_1m_2}\left(U''-\frac{U'}{r}\right)\left[\bm S^2-3\frac{(\bm S\cdot\bm r)^2}{r^2}\right]\right\},
\end{eqnarray}
\begin{eqnarray}\label{sggpot}
    V^S_{gg}&=&{\cal C}_5 \alpha_S\left\{U-\frac{1}{3m^2}\left(3-\bm S^2\right)\bm\nabla^2U-\frac{1}{2m^2}\left[\bm p_1U\bm p_2-(\bm p_1\cdot\bm r)\frac{U'}{r}(\bm r\cdot\bm p_2)\right]\right.\nonumber\\
    &&+\frac{3}{2m^2}\bm L\cdot\bm S\ \frac{U'}{r}-\frac{1}{4m^2}\bm \delta\cdot\bm \Omega\ \frac{U'}{r}\left.+\frac{1}{6m^2}\left(U''-\frac{U'}{r}\right)\left[\bm S^2-3\frac{(\bm S\cdot\bm r)^2}{r^2}\right]\right\},
\end{eqnarray}
\begin{eqnarray}\label{sqgpot}
        V^S_{qg} &=&\, {\cal C}_7\alpha_S\left\{U-\frac{1}{2m_gm_q}\left[\bm p_1U\bm p_2-(\bm p_1\cdot\bm r)\frac{U'}{r}(\bm r\cdot\bm p_2)\right]\right.\nonumber\\
        &&-\left(\frac{11}{12m_gm_q}-\frac{1}{6m^2_g}-\frac{1}{8m^2_q}-\frac{1}{3m_gm_q}\bm S^2\right)\bm\nabla^2U\left(\frac{m^2_q+m^2_g+4m_gm_q}{4m_g^2m_q^2}\right)\bm L\cdot\bm S \ \frac{U'}{r}\nonumber\\
        &&-\frac{1}{4m_gm_q}\bm \delta\cdot\bm \Omega\ \frac{U'}{r}-\left(\frac{m_g^2-m_q^2}{4m_q^2m_g^2}\right)\bm L\cdot\bm \delta\ \frac{U'}{r}-\frac{(m_q-m_g)}{4(m_g+m_q)m_gm_q}\bm S\cdot\bm \Omega\ \frac{U'}{r}\nonumber\\
    &&+\frac{1}{6m_gm_q}\left(U''-\frac{U'}{r}\right)\left[\bm S^2-3\frac{(\bm S\cdot\bm r)^2}{r^2}\right]\left. -\frac{m_q-m_g}{2m^2_gm_q}\left(U''-\frac{U'}{r}\right)\frac{\bm r Q\bm r}{r^2}\right\}.
\end{eqnarray}
The symmetrizations on the noncommuting operators have not been explicitly written in order to clarify to formula. It appears that all the scattering potentials we compute almost share the same structure, the only difference coming in the coefficients of $\bm\nabla^2 U$, typically responsible for the contact interaction. The quadrupolar term is trivially zero except for the (anti)quark-gluon interactions. In this last case, we denoted the gluon and the (anti)quark masses by $m_g$ and $m_q$ respectively, in order to avoid confusion. It is worth mentioning that, if $A_{ij}$ is the coefficient of $\bm\nabla^2 U$ in the scattering potential $V^S_{ij}$, it can be checked that 
\begin{eqnarray}
A_{qq}&=&-\frac{1}{8m^2_1}-\frac{1}{8m^2_2}-\frac{2\bm S_1\cdot\bm S_2}{3m_1m_2},\\
A_{gg}&=&-\frac{1}{6m^2}-\frac{1}{6m^2}-\frac{2\bm S_1\cdot\bm S_2}{3m^2},\\
A_{qg}&=&-\frac{1}{6m^2_g}-\frac{1}{8m^2_q}-\frac{2\bm S_q\cdot\bm S_g}{3m_gm_q}.
\end{eqnarray}
It appears obviously that the differences only depend on the nature of the interacting particles, each quark bringing a contribution $-1/8m^2_q$, while this contribution is equal to $-1/6m^2_g$ for a gluon.

Clearly, the Coulomb form of $U$ comes from the propagator of a massless particle. Following ref.~\cite{bla}, the expressions (\ref{sqqpot}), (\ref{sggpot}), and (\ref{sqgpot}) are valid for any $U(r)$, this function being the Fourier transform of the propagator of an exchanged vector particle. Thus, for a massive gluon, one should simply replace $U(r)$ by a Yukawa potential $\tilde U(r)$. Moreover, following eq. (\ref{green}), $\bm\nabla^2\tilde U=m^2\tilde U-4\pi\delta^3(\bm r)$.

Various gluon-gluon potentials involving massive gluons can be found in the literature (see for example refs.~\cite{corn,hou}), but they are neither equal between each other, nor equivalent to our potential. In recent works \cite{Koma:2005nq}, the coefficients of the different terms ($\bm{ L\cdot S}$, $\bm{ L\cdot\delta}$, ...) of the potential between two heavy quarks was found numerically by lattice QCD computations. These fitted coefficients agree with the Fermi-Breit potential. It could be interesting to find, by the same procedure, the same coefficients for two static sources in a color octet. This could lead to identify the most relevant gluon-gluon potential. Moreover, an accurate computation of various hadrons spectra will be able to check the relevance of one approach or another, through a comparison with experimental data or with lattice QCD.  

\section{Validity of the approach}\label{valid}

\subsection{The strong coupling constant}

The effective potentials we computed up to now involve Feynman diagrams of order $g^2$, without loop. It is well known however that the inclusion of the one-loop diagrams leads to the conclusion that the strong coupling constant depends on the exchanged momentum through
\begin{equation}\label{asr}
	\alpha_S(q^2)=\frac{12\pi}{(33-2N_f)\ln\left(\frac{q^2}{\Lambda^2}\right)},
\end{equation}
where $N_f$ is the number of quark flavors whose masses are lower than $q^2$, and where $\Lambda$ is the famous lambda-QCD parameter, whose value is around $300$ MeV. Clearly, $\alpha_S(q^2)$ is smaller and smaller for increasing $q^2$, QCD becoming perturbative at high momentum. As an illustration, it is nowadays accepted that $\alpha_S(m^2_Z)=0.1176\pm0.0020$ \cite{PDG}, with $m_Z\approx 92$~GeV. For $q^2\approx\Lambda^2$ however, QCD is nonperturbative at low momentum since the strong coupling constant~(\ref{asr}) blows up. This low-momentum domain precisely concerns the bound states physics we are interested in. Consequently, it appears necessary to wonder whether the potentials we obtain are valid or not to describe hadronic bound states. 

Lattice QCD clearly shows that the static potential between a quark and an antiquark is nicely fitted by a Cornell shape, that is $ar-4\alpha_S/3r$, for $a\approx0.20$ GeV$^2$ and $\alpha_S\approx0.22$ \cite[p. 42]{balirep}. Such a form suggests that the total energy between the quarks can be roughly separated into a confining (nonperturbative) part at long-range ($ar$), and a ``residual", short-range, Coulombic part ($-4\alpha_S/3r$). In this picture, $\alpha_S$ should be interpreted as a small, effective, strong coupling constant. We point out that, although the Cornell potential correctly matches the data, it should only be seen as an interesting approximation of the real static potential, for example because it neglects the interplay between perturbative and nonperturbative effects at intermediate distances. Moreover, the calculations of ref.~\cite{pineda} forbid the presence of a linear term at short distances. But, the Cornell potential is clearly the simplest way to interpolate between a Coulomb potential at short distances and a linear confining one at long distances, both asymptotic behaviors being confirmed by theoretical studies~\cite{weis,pineda}. Anyway, the short-range part of the static potential appears to be mostly given by a perturbative Coulomb term, i.e. the static approximation of all the potential we computed up to now. Let us note that strong arguments indicate that this picture is also valid not only for bound states of heavy quarks and antiquarks, but also for bound states of massless gluons~\cite{glue2}, thus in the light particle sector. 

Several attempts have been made in order to compute $\alpha_S(q^2)$ at all orders, especially by solving the Dyson-Schwinger equations (see a rather complete list of references in ref.~\cite{free}). All these approaches qualitatively lead to
\begin{equation}\label{asr2}
	\alpha_S(q^2)=\frac{12\pi}{(33-2N_f)\ln\left(\frac{q^2+\xi^2(q^2)}{\Lambda^2}\right)},
\end{equation}
where $\xi(q^2)$ is a monotonic function such that $\xi^2(0)>0$ and $\xi^2(q^2\rightarrow\infty)\rightarrow0$. However, the explicit formula giving $\xi(q^2)$ is different following the different works. Equation~(\ref{asr2}) states that the strong coupling constant remains finite for $q^2=\Lambda^2$, and tends to a maximal value $\alpha_S(0)$ -- one speaks of ``freezing" of the strong coupling constant. The value of $\alpha_S(0)$ is still a matter of research, but no theoretical argument forbids $\alpha_S(0)<1$, and some studies even allow for $\alpha_S(0)$ to be as low as $0.4$~\cite{free}, just around the typical values which are used in potential models~\cite{expe}. Such a freezing scenario is coherent with the perturbative point of view that we adopted, and justifies in particular the assumption that the lowest order Feynman diagrams already give a relevant picture of the short-range interactions between two confined particles. 

For completeness, it should be pointed out that, although providing a coherent framework to understand to success of potential models and how to build them, the freezing of the strong coupling constant is not universally accepted, see for example the lattice computations of ref.~\cite{mason} in which no freezing is observed.    

\subsection{Nonrelativistic expansion}
Another important ingredient of our computations is a nonrelativistic expansion of the ``wave functions" (Dirac spinors and polarization vectors) in powers of $\bm p^2/\mu^2$. Such an expansion is clearly valid for heavy quarks since $\left\langle \bm p^2/\mu^2\right\rangle$ $\approx \left\langle \bm p^2/m^2\right\rangle\ll1$. In this last case, our quark-quark potential reduces to the usual Fermi-Breit one, which is widely used in potential models. 

For light quarks and massless constituent gluons, it is readily checked that $\left\langle \bm p^2/\mu^2\right\rangle \approx 1$, and the nonrelativistic expansion is formally no longer valid. However, although no theoretical argument is able to justify it, the extension of the Fermi-Breit potential to light mesons and baryons has been shown to reproduce the experimental data for a long time with a quite good agreement~\cite{old,old2,expe}. Even the glueball spectrum, computed within a potential model where we used the gluon-gluon potential that we obtained in this work, is in agreement with lattice QCD when massless constituent gluons are used~\cite{expe}. By doing explicit computations, one is thus led to the empirical conclusion that the first relativistic corrections in $\bm p^2/\mu^2$ contain most of the physically relevant information. But, it is clear that light quarks are at the limit of the validity range of potential approaches.

For what concerns the constituent gluons, it is worth mentioning that, since the work of ref.~\cite{corn}, several studies have given theoretical arguments favoring massive constituent gluons, with a mass which could be as high as $m_g=0.8$ GeV. In this case, our gluon-gluon potentials remain valid, but with $U(r)={\rm e}^{-m_gr}/r$, as we argued in Sec.~\ref{applic}. At such high masses, $\left\langle \bm p^2/m^2_g\right\rangle$ becomes small enough for the nonrelativistic expansion to be justified. The introduction of massive constituent gluons, in gluon-gluon as well as in (anti)quark-gluon potentials, could then bypass the conceptual problems caused by the nonrelativistic expansion.

\section{Color factors}\label{colfact}
\subsection{Clebsh-Gordan coefficients}\label{cgc}
We have now to compute the value of the various color operators ${\cal O}_{n;A'B'AB}$ that we introduced in the previous sections. Up to now, we did not impose the two interacting particles to be in a particular representation of $SU(3)$. Let us illustrate this with the $qq$ interaction in a baryon. The color functions of the quarks were of course chosen to be in the fundamental representation of $SU(3)$, but we did not explicitly asked that the $qq$ pair had to be in the conjugate representation in order to make a singlet when coupled to the third quark. To include this point in our discussion, we have to contract the color operators with the appropriate Clebsh-Gordan coefficients (CGC) of the \textit{in} and \textit{out} states. Let $A,B$ be the generic color indices of the two incoming particles, and $A',B'$ those of the outcoming particles. In all our potentials, ${\cal O}_{n;A'B'AB}$ has thus to be replaced by
\begin{equation}\label{ccomp}
    {\cal C}_n=\sum_{ABA'B'} F^{*\, A'B'}_C {\cal O}_{n;A'B'AB} F^{AB}_C,
\end{equation}
where $F^{AB}_C$ are the CGC. $C$ stands for the color index of the coupling, which is the same for the in and out states (the color is conserved during the interaction). In every cases, one can check that the result is independent of $C$.
\par Let us now compute the $F^{AB}_C$, in a similar way as it is done it ref.~\cite[p. 51]{car}. Let $\phi^{\bm X}$ be the color wave function of a particle in the representation $\bm X$ of $SU(3)$. In other words, $\delta_{\vec\epsilon}\, \phi^{\bm X} = i\vec\epsilon\cdot\vec T^{\bm X} \phi^{\bm X}$ with $\vec T^{\bm X}$ the generators of the representation ${\bm X}$ and $\vec\epsilon$ defining an infinitesimal gauge transformation (in this section, the vectors will be denoted with an arrow for convenience). Because the $F^{AB}_C$ are general CGC that couple the representations $\bm X$ and $\bm Y$ to form the representation $\bm Z$, we can find them by requiring
\begin{equation}\label{cg}
    \delta_{\vec\epsilon}\, \left[\phi^{\bm X}_A F^{AB}_C \phi^{\bm Y}_B\right]= i\vec\epsilon\cdot(\vec{T}^{\bm Z})^{\ D}_C \left[\phi^{\bm X}_AF^{AB}_D\phi^{\bm Y}_B\right].
\end{equation}
Moreover, the CGC computed thanks to eq.~(\ref{cg}) have to be normalised with the relation
\begin{equation}
    \sum_{AB} F^{*\, AB}_{C}F^{AB}_{D} = \delta_{CD}.
\end{equation}
Our method can be in principle applied to any representation of $SU(3)$. But in this work, we will restrict ourselves to systems involving only the representations $\bm1$, $\bm3$, $\bar{\bm3}$ and $\bm8$, which are particularly relevant for hadron spectroscopy. In particular, the quarks, antiquarks and gluons are in the $\bm3$, $\bar{\bm3}$, and $\bm8$ representations. Some CGC are computed in table~\ref{tab:cgc}. As it is shown in the second column, they can be used to describe mesons, baryons, glueballs, hybrids, etc. We recall that $\varepsilon^{ij}_k$ is the Levi-Civita symbol, and that the constants $f^{ab}_c$ and $d^{ab}_c$ are defined by the well-known relations
\begin{equation}
\left[\lambda^a,\lambda^b\right]=2i \sum_c f^{ab}_c \lambda^c,
\end{equation}
\begin{equation}
    \left\{\lambda^a,\lambda^b\right\}=2\sum_c d^{ab}_c \lambda^c+\frac{4}{3}\delta^{ab}.
\end{equation}
\begin{table}[hb] \caption{Normalised CGC for two particles in the respective representations $\bm X$ and $\bm Y$ of $SU(3)$, coupled so that they are in the representation $\bm Z$. The indices $i,j,k$ take values from $1$ to $3$, and $a,b,c$ from $1$ to $8$. The color indices of the two particles are the upper indices. The lower one is the color of their coupling.}
\begin{center}
        \begin{tabular}{lll}
        \hline
    $\left[\bm X,\bm Y\right]^{\bm Z}$ & Example & CGC\\
    \hline
$\left[\bm3,\bm3\right]^{\bar{\bm{3}}}$  & $qq$ in $qqq$ & $\frac{1}{\sqrt{2}}\varepsilon^{ij}_{\ \;k}$ \\
$\left[\bar{\bm3},\bar{\bm3}\right]^{\bm{3}}$  & $\bar q\bar q$ in $\bar q\bar q \bar q$  & $\frac{1}{\sqrt{2}}\varepsilon^{ij}_{\ \;k}$ \\
$\left[\bm3,\bar{\bm{3}}\right]^{\bm1}$  & $q\bar q$ in $q\bar q$ &     $\frac{1}{\sqrt{3}}\delta^{ij}$ \\
$\left[\bm3,\bar{\bm{3}}\right]^{\bm8}$  & $q\bar q$ in $q\bar qg$ & $\frac{1}{\sqrt{2}}(\lambda_a)^{ij} $\\
$\left[\bm3,\bm8\right]^{\bm3}$          & $qg$ in $q\bar qg$ &     $\frac{\sqrt{3}}{4}(\lambda^a)^i_{\ j}$\\
$\left[\bar{\bm3},\bm8\right]^{\bar{\bm3}}$ & $\bar q g$ in $q\bar qg$ & $\frac{\sqrt{3}}{4}(\lambda^a)_i^{\ j}$\\
$\left[\bm8,\bm8\right]^{\bm1}$          & $gg$ in  $gg$  &  $ \frac{1}{\sqrt{8}}\delta^{ab}$\\
$\left[\bm8,\bm8\right]^{\bm8_s}$        & $gg$ in  $ggg$  & $\frac{1}{\sqrt{3}}f^{ab}_{\ \ \;c}$\\
$\left[\bm8,\bm8\right]^{\bm8_a}$        & $gg$ in $ggg$  & $\sqrt{\frac{3}{5}}d^{ab}_{\ \ \;c}$\\
\hline
        \end{tabular}
\end{center}
    \label{tab:cgc}
\end{table}

\subsection{Computation of the color factors}\label{colfactc}

We first consider the color operators which can appear in the quark-quark diagrams of sec.~\ref{qqint}. They read
\begin{subequations}\label{color:qq}
\begin{eqnarray}
{\cal O}_{1;ijkl} &=& \frac{1}{4}(\lambda^a)_{ji}(\lambda_a)_{lk}\equiv \frac{\vec\lambda_{1}}{2}\cdot\frac{\vec\lambda_{2}}{2},\\
{\cal O}_{2;ijkl} &=& -\frac{1}{4}(\lambda^a)_{ji}(\lambda_a)_{kl} \equiv \frac{\vec\lambda_{1}}{2}\cdot\left[-\frac{\vec\lambda_{2}^*}{2}\right], \\
{\cal O}_{3;ijkl} &=&\frac{1}{4}(\lambda^a)_{ij}(\lambda_a)_{kl}\equiv
\left[-\frac{\vec\lambda_{1}^
*}{2}\right]\cdot\left[-\frac{\vec\lambda_{2}^*}{2}\right],\\
{\cal O}_{4;ijkl} &=& \frac{1}{4}(\lambda^a)_{ki}(\lambda_a)_{jl}\equiv \frac{4}{9}+ \frac{1}{3}\frac{\vec\lambda_{1}}{2}\cdot\left[-\frac{\vec\lambda_{2}^*}{2}\right],
\end{eqnarray}
\end{subequations}
where the index in $\vec\lambda_i$ labels the two interacting particles.
The first three operators are present in the scattering diagram of two quarks, a quark and an antiquark, and two antiquarks respectively. The fourth color operator appears in the quark-antiquark annihilation diagram. It is logically a projector on the octet state, since this annihilation can not occur if the quark and the antiquark are in a singlet. The color factors corresponding to the operators (\ref{color:qq}) are summed up in table~\ref{Tab:color:qq}.

\begin{table}\caption{Values of the color factors corresponding to the operators \eqref{color:qq} for the different couplings of table~\ref{tab:cgc}. The ${\cal C}_j$ are computed thanks to formula~(\ref{ccomp}) in the relevant cases. }
\begin{center}
\begin{tabular}{ccccc}
\hline
$\left[\bm X,\bm Y\right]^{\bm Z}$ & ${\cal C}_1$ & ${\cal C}_2$ & ${\cal C}_3$ & ${\cal C}_4$\\
 \hline
$\left[\bm3,\bm3\right]^{\bar{\bm{3}}}$ & $-2/3$ & $\times$ & $\times$ & $\times$\\
$\left[\bar{\bm3},\bar{\bm3}\right]^{\bm{3}}$ & $\times$ & $\times$& $-2/3$ & $\times$ \\
$\left[\bm3,\bar{\bm{3}}\right]^{\bm1}$ & $\times$ & $-4/3$ & $\times$ & $0$\\
$\left[\bm3,\bar{\bm{3}}\right]^{\bm8}$ & $\times$ & $1/6$ &$\times$ & $1/2$ \\
\hline
\end{tabular}
\label{Tab:color:qq}
\end{center}
\end{table}

\par If we denote $\vec T^{\bm X}_i$ as the generators of $SU(3)$ in the representation corresponding to the particle $i$, we can check that the relation
\begin{equation} \label{Cas}
    \vec T^{\bm X}_1\cdot\vec T^{\bm Y}_2 = \frac{1}{2}\left[(\vec T^{\bm Z}_{12})^{\, 2}-\vec T^{\bm X\, 2}_1-\vec T_2^{\bm Y\, 2}\right]
\end{equation}
is verified as expected for the scattering diagrams. $\vec T^{\bm Z}_{12}$ are the generators for the representation of the bound state of $1$ and $2$. Let us note that, in eq.~(\ref{color:qq}), $\vec T^{\bm 3}=\vec\lambda/2$ and $\vec T^{\bar{\bm 3}}=-\vec\lambda^*/2$.

\par The interactions between two gluons are treated in sec.~\ref{ggint}. It appears that, among the four diagrams, there are only two independent color operators
\begin{subequations}\label{color:gg}
\begin{eqnarray}
{\cal O}_{5;abcd} &=&  -f_{\ ba}^ef_{edc},\\
{\cal O}_{6;abcd} &=&   f_{\ ca}^ef_{ebd}.
\end{eqnarray}
\end{subequations}
The first one appears in the scattering of two gluons, while the annihilation diagram is proportional to second one. The corresponding color factors are given in table~\ref{Tab:color:gg}. Our results agree with those of ref.~\cite{hou}. We can observe that the contribution of the annihilation diagram is zero when the two gluons are in a singlet, as expected, but also when they are in a symmetric octet. This can be encountered in a glueball formed of three gluons, and a posteriori justified by observing that the effective potential for the annihilation diagram is proportional to $\delta^3(\bm r)$. Thus, its contribution is nonzero only for the $s$-wave, which has an symmetric space wave function and an antisymmetric spin wave function. In order for the total wave function to be symmetric (we are dealing with two identical bosons), the color function needs thus to be antisymmetric. That is why the annihilation can not occur in the symmetric octet. Let us also note that eq.~(\ref{Cas}) can be used in the scattering diagram of two gluons, but in this case, $(T^{\bm 8\, a})_{bc}=-if_{\ bc }^a$ have to be used, as the generators of the $\bm 8$ representation.
\begin{table}\caption{Same as table~\ref{Tab:color:qq}, but for the color operators \eqref{color:gg}.}
\begin{center}
\begin{tabular}{lcc}
\hline
$\left[\bm X,\bm Y\right]^{\bm Z}$ & ${\cal C}_5$ & ${\cal C}_6$ \\
 \hline
$\left[\bm8,\bm8\right]^{\bm1}$ & $-3$ & $0$ \\
$\left[\bm8,\bm8\right]^{\bm8_s}$  & $-3/2$ & $0$ \\
$\left[\bm8,\bm8\right]^{\bm8_a}$  & $-3/2$ & $-3$ \\
\hline
\end{tabular}
\label{Tab:color:gg}
\end{center}
\end{table}

\par The last case concerns the quark-gluon interactions. Five different color operators appear in the diagrams
\begin{subequations}\label{color:qg}
\begin{eqnarray}
{\cal O}_{7;bcjk} &=& (\frac{\lambda^a}{2})_{kj}(-if_{acb}) ,\\
{\cal O}_{8;bcjk} &=& \frac{1}{4} (\lambda^c\lambda^b)_{kj}, \\
{\cal O}_{9;bcjk} &=& \frac{1}{4} (\lambda^b\lambda^c)_{jk}, \\
{\cal O}_{10;bcjk} &=& \frac{1}{4} (\lambda^b\lambda^c)_{kj},\\
{\cal O}_{11;bcjk} &=&\frac{1}{4} (\lambda^c\lambda^b)_{jk} ,
\end{eqnarray}
\end{subequations}
and the corresponding color factors are listed in table~\ref{Tab:color:qg}. Again, ${\cal C}_7$ can be computed with eq.~(\ref{Cas}), in agreement with ref.~\cite{Mand}.
\begin{table}\caption{Same as table~\ref{Tab:color:qq}, but for the color operators \eqref{color:qg}.}
\begin{center}
\begin{tabular}{cccccc}
\hline
$\left[\bm X,\bm Y\right]^{\bm Z}$ & ${\cal C}_7$ & ${\cal C}_8$ & ${\cal C}_9$ & ${\cal C}_{10}$ & ${\cal C}_{11}$\\
 \hline
$\left[\bm3,\bm8\right]^{\bm3}$ & $-3/2$ & $4/3$ & $\times$ & $-1/6$ & $\times$ \\
$\left[\bar{\bm3},\bm8\right]^{\bar{\bm3}}$  & $-3/2$ & $\times$ &  $4/3$ & $\times$ & $-1/6$ \\
\hline
\end{tabular}
\label{Tab:color:qg}
\end{center}
\end{table}

\section{Conclusion and outlook}\label{conclu}
We have computed in this paper the effective potentials arising from second order Feynman diagrams involving quarks, antiquarks, or gluons. Our expressions remain defined for the lightest particles, because the masses appearing in our potentials are seen as the constituent ones, not the bare ones. For instance, in the nucleon, the constituent mass of the $u,d$ quarks are about $300$ MeV. In glueballs, the typical value for the gluon constituent mass is $500$ MeV \cite{Math}. It is worth mentioning that a recent comparison between quark models and large $N_c$ expansion of QCD also supports the development of the relativistic corrections in powers of $1/\mu$~\cite{lnc}.

We firstly re-obtained the well-known Fermi-Breit potential for two interacting quarks. Then, we computed the effective potential between two constituent gluons, and we derived an effective potential for the quark-gluon interaction and showed that in this case a quadrupolar interaction is present, due to the asymmetry of the system. Our results are valid at the order $c^{-2}$, which ensures us to have correct hyperfine terms, like spin-orbit, tensor force,\, \dots

As some studies argue that a nonzero mass must be confered to the exchanged gluons, we showed how to include in a simple way such a mass starting from our expressions. Whether the exchanged gluons are massive or not clearly leads to very different potentials. The predictions of these two approaches could be compared, for example to the glueball spectra obtained in lattice QCD, in order to check their relevance. However, we stress that the goal of this work is not to decide whether the gluons are massive or not: We computed general effective potentials in a systematic way from QCD, and were led to expressions which remain valid for light particles. Furthermore, for the first time, we obtained a quark-gluon interaction potential. 

\par As an important check of the framework that we presented here, we can mention that we recently used it to compute light and heavy meson as well as two-gluon glueball spectra~\cite{expe}. It can be seen in this last reference that the agreement between experimental data and our theoretical spectra is very satisfactory, and that the glueball spectrum we get is mostly located in the error bars of the lattice QCD one. Consequently, we think that the various potentials which are presented in this study can be successfully used as short-range terms within potential models describing not only usual hadrons, but also exotics like two or three-gluon glueballs, and hybrid $q\bar q g$ mesons, which currently deserve much interest. Motivated by the results of ref.~\cite{expe}, we think that our results can be seen as a basis for further computations of exotic hadron spectra, that we leave for future works.

The authors thank Dr. Claude Semay for many fruitful and stimulating discussions. 

\begin{appendix}

\section{Feyman rules for QCD}\label{feyn}
We recall in this section the Feynman rules we use. They are identical to those of ref.~\cite[p. 383]{Ynd} up to some constant normalisation factors, which were changed to obtained more suitable expressions for our calculations. For the \textit{in} and \textit{out} states wave functions, we have:
%
%
%
%
%
\begin{figure}[ht]
  \begin{center}
    \includegraphics*[width=8cm]{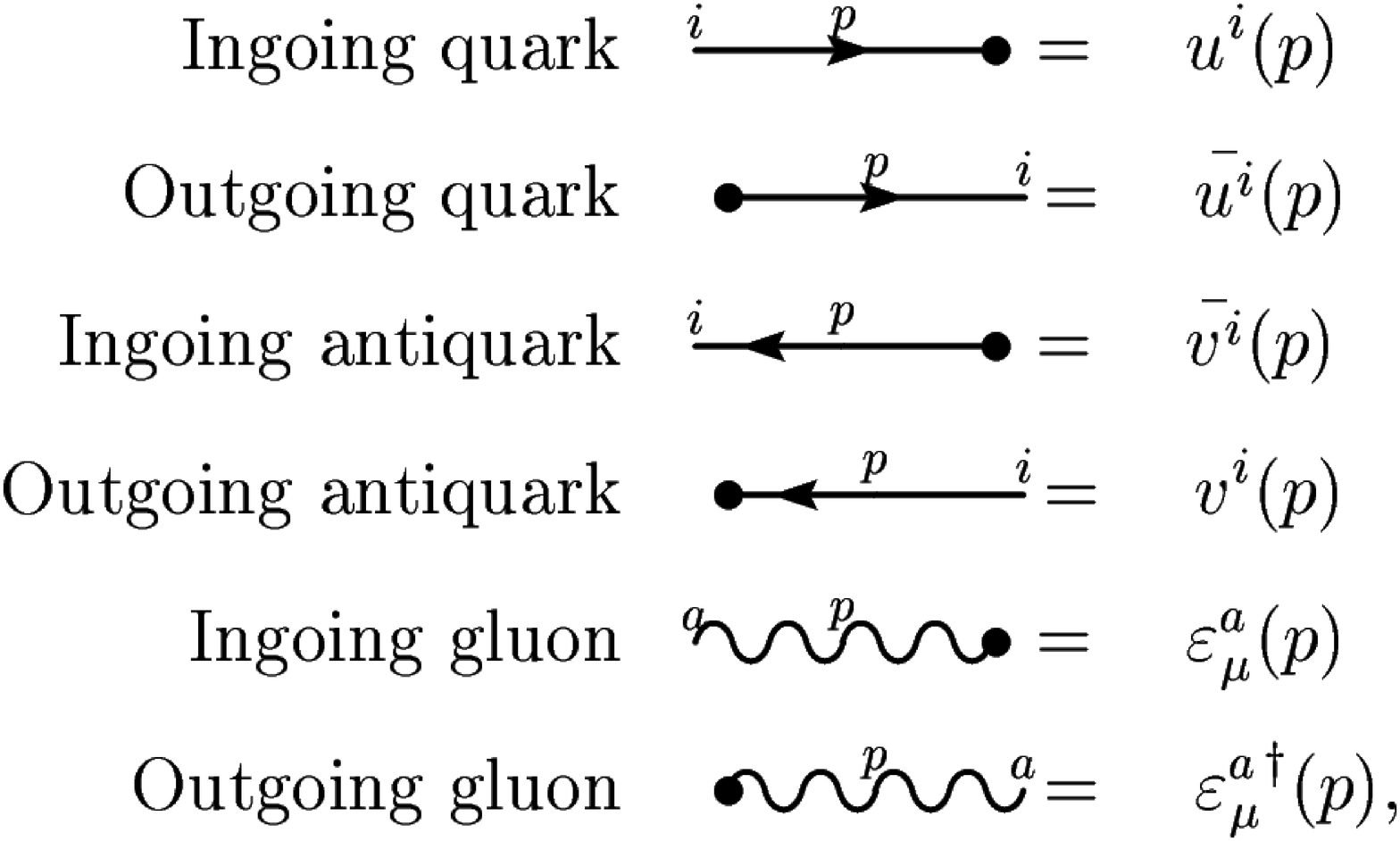}
  \end{center}
  \label{feyn0}
\end{figure}

\noindent where $p$ is the $4-$momentum of the particle. As color indices, we used $a=1,...,8$ for a particle in the adjoint representation, and $i=1,2,3$ for a particle in the fundamental representation.

The quark propagator and the gluon propagator in Lorentz
gauge read

%
\begin{figure}[ht]
  \begin{center}
    \includegraphics*[width=6cm]{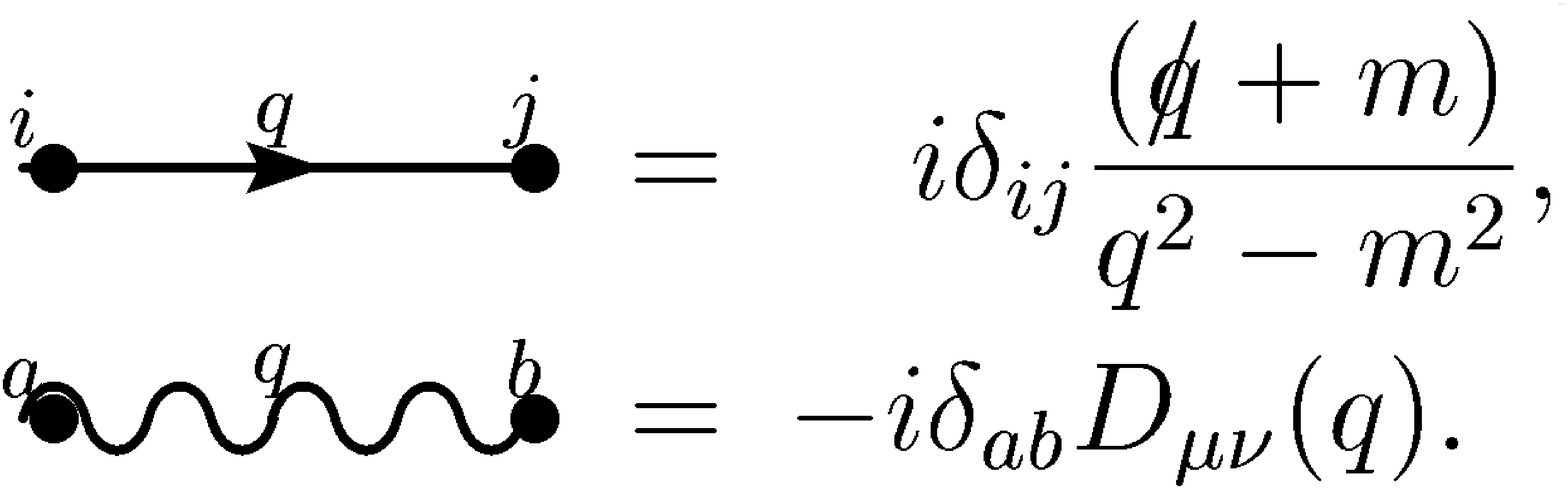}
  \end{center}
  \label{feyn1}
\end{figure}

\noindent $D_{\mu\nu}(q)$ is defined as
\begin{equation}\label{Ddef}
D_{\mu\nu}(q)=\dfrac{1}{q^2}\left[\eta_{\mu\nu}+(\theta-1)\dfrac{q_\mu
q_\nu}{q^2}\right],
\end{equation}
where $\theta$ is the gauge parameter, and $\eta={\rm diag}(+---)$ is the Minkowski metric. $\mu,\nu=0,...,3$ are used as space-time indices. As usual, $q\!\!\!/=\gamma^\mu q_\mu$, with $\gamma^\mu$ the Dirac matrices, satisfying
\begin{equation}
    \left\{\gamma^\mu,\gamma^\nu\right\}=2 \eta^{\mu\nu}.
\end{equation}
These matrices can be explicitly represented by
\begin{equation}
\bm \gamma^0=\begin{pmatrix}\mathun&0\\0&-\mathun\end{pmatrix},\quad    \bm\gamma=\begin{pmatrix}0&\bm\sigma\\-\bm\sigma&0\end{pmatrix},
\end{equation}
with ${\bm \sigma}$ the Pauli matrices.

We also need the contributions of the different vertices. They are given by
%
%
\begin{figure}[ht]
  \begin{center}
    \includegraphics*[width=10.0cm]{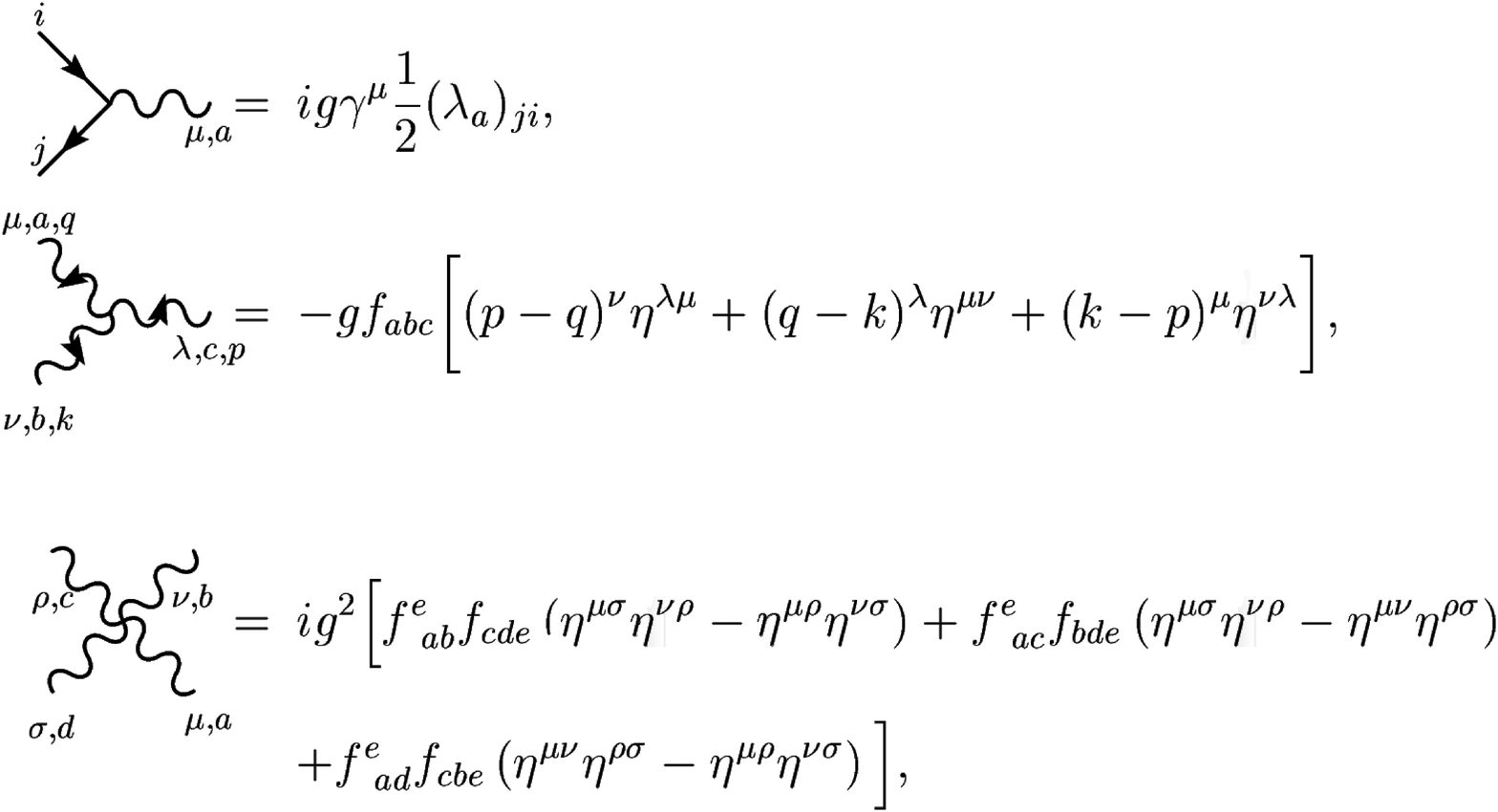}
  \end{center}
  \label{feyn2}
\end{figure}

\noindent where $\lambda_a$ are the Gell-Mann matrices, and $f_{abc}$ are the structure constants of $SU(3)$. Finally, for each matrix element, we have to add a factor $(2\pi)^4 \delta^4(P-P')$ for the energy-momentum conservation. $P$ and $P'$ are the initial and final $4$-momentum respectively.
\section{Some Fourier transforms}\label{fourier}
We present in this section useful Fourier transforms that are taken from ref.~\cite[p. 282]{land}. We define the Fourier transform of a function $g(\bm q)$ as
\begin{equation}\label{def:fourier}
   {\cal F}\left[g(\bm q)\right]= \int\frac{d^3\bm q}{(2\pi)^3}\;e^{-i\bm{q\cdot
    r}}g(\bm q),
\end{equation}
where the bold quantities always denote a vector. 

Then, one can check that the following equalities hold
\begin{eqnarray}
{\cal F}\left(1\right)&=& \delta^3(\bm r), \\
{\cal F}\left[\frac{4\pi}{\bm q^2+m^2}\right]&=& \frac{{\rm e}^{-mr}}{r}\quad m\geq0, \\
{\cal F}\left[\frac{4\pi(\bm{a\cdot q})(\bm{b\cdot q})}{
    \bm q^4}\right]&=& \frac{1}{2r}\left[\bm{a\cdot
    b}-\frac{(\bm{a\cdot r})(\bm{b\cdot r})}{r^2}\right],\\
{\cal F}\left[\frac{4\pi(\bm{a\cdot q})(\bm{b\cdot q})}{
    \bm q^2+m^2}\right]&=& -\left[(\bm{a\cdot\nabla})(\bm{b\cdot\nabla})+\frac{1}{3}(\bm{a\cdot b})\bm\nabla^2 \right] \frac{{\rm e}^{-mr}}{r},\label{eqbiz}\\
{\cal F}\left[\frac{4\pi i\bm{a\cdot}(\bm q\times\bm
    p)}{\bm q^2+m^2}\right]&=&-[\bm{a\cdot}(\bm r\times\bm p)]\frac{1}{r}\frac{\partial}{\partial r}\frac{{\rm e}^{-mr}}{r},
\end{eqnarray}
with $\bm a$ and $\bm b$ vectors or matrix vectors. All these identities hold for massless propagators, but the limit $m=0$ has to be taken before applying the derivatives. Let us recall the relation
\begin{equation}\label{green}
    (\bm\nabla^2-m^2)\frac{{\rm e}^{-mr}}{r}=-4\pi\delta^3(\bm r),
\end{equation}
which is useful in the computation of eq.~(\ref{eqbiz}).
\end{appendix}

\end{document}